\documentclass[%
reprint,
superscriptaddress,
amsmath,amssymb,
aps,
prl,
]{revtex4-2}

\usepackage{dcolumn}

\usepackage[T1]{fontenc}
\usepackage[utf8]{inputenc}
\usepackage{bbold}
\usepackage{amsmath}
\usepackage{amssymb}
\usepackage{graphicx}
\usepackage{graphicx}
\usepackage{subscript}
\usepackage[unicode=true,pdfusetitle,
bookmarks=true,bookmarksnumbered=false,bookmarksopen=false,
breaklinks=false,pdfborder={0 0 1},backref=false,colorlinks=false]
{hyperref}
\usepackage{xcolor}
\usepackage[title]{appendix}

\usepackage{bm}
\usepackage{physics}
\usepackage{mhchem}
\usepackage{textcomp}
\usepackage{indentfirst}
\usepackage{listings}
\usepackage{svg}
\usepackage[export]{adjustbox}
\usepackage{lipsum} 

\newcommand{\lyxmathsym}[1]{\ifmmode\begingroup\def\b@ld{bold}
	\text{\ifx\math@version\b@ld\bfseries\fi#1}\endgroup\else#1\fi}

\begin{document}
	\title{Excitonic optical absorption in strained monolayer \ce{CrSBr}}
	
	\author{Maur\'icio F. C. Martins Quintela}
	\email{mfcmquintela@gmail.com, corresponding author}
	\affiliation{Condensed Matter Physics Center (IFIMAC), Universidad Aut\'{o}noma de Madrid, E-28049 Madrid, Spain}
	\affiliation{Departamento de F\'isica de la Materia Condensada, Universidad Aut\'{o}noma de Madrid, E-28049 Madrid, Spain}
	
	\author{Guilherme J. Inacio}
	\affiliation{Departamento de F\'isica de la Materia Condensada, Universidad Aut\'{o}noma de Madrid, E-28049 Madrid, Spain}
	
	\author{Miguel S\'a}
	\affiliation{Departamento de F\'isica de la Materia Condensada, Universidad Aut\'{o}noma de Madrid, E-28049 Madrid, Spain}
	
	\author{Giovanni Cistaro}
	\affiliation{Theory and Simulation of Materials (THEOS), \'Ecole Polytechnique F\'ed\'erale de Lausanne (EPFL), CH-1015, Lausanne, Switzerland}
	
	\author{Alberto M. Ruiz}
	\affiliation{Instituto de Ciencia Molecular, Universitat de Val\`encia, E-46980 Paterna, Spain}
	
	\author{Jos\'e J. Baldov\'i}
	\affiliation{Instituto de Ciencia Molecular, Universitat de Val\`encia, E-46980 Paterna, Spain}
	
	\author{Juan J. Palacios}
	\affiliation{Condensed Matter Physics Center (IFIMAC), Universidad Aut\'{o}noma de Madrid, E-28049 Madrid, Spain}
	\affiliation{Departamento de F\'isica de la Materia Condensada, Universidad Aut\'{o}noma de Madrid, E-28049 Madrid, Spain}
	\affiliation{Instituto Nicol\'as Cabrera (INC), Universidad Aut\'onoma de Madrid, E-28049 Madrid, Spain}
	
	\author{Antonio Pic\'on}
	\email{antonio.picon@csic.es, corresponding author}
	\affiliation{Condensed Matter Physics Center (IFIMAC), Universidad Aut\'{o}noma de Madrid, E-28049 Madrid, Spain}
	\affiliation{Instituto de Ciencia de Materiales de Madrid (ICMM-CSIC), E-28049 Madrid, Spain}
	
	\begin{abstract}
		
		Recently, the isolation of 2D magnetic materials has opened several avenues for possible new applications in spintronics. Among these materials, \ce{CrSBr} has sparked interest due to its relatively high Curie temperature, highly anisotropic lattice structure, and high structural stability. These properties ran along others shared by any atomically thin material such as its outstanding deformation capacity and a strong optical response dominated by excitonic effects. The combination of these properties provides a fairly uncharted playground where to explore the interplay between magnetism and optical excitations. Here, we focus our attention on the theoretical optical response of \ce{CrSBr} under several distinct strain configurations, analyzing the resulting changes to both the excitonic peaks and overall shape of the diagonal components of the linear conductivity tensor.
	\end{abstract}
	
	\maketitle

	\section{\label{sec:intro}Introduction}
	
	Ever since the first isolation of two-dimensional (2D) or atomically thin materials \cite{doi:10.1126/science.1102896}, it has been understood that excitons are key to the understanding of the sub-gap optical response when these materials  are semiconducting or insulating \cite{PhysRevLett.105.136805,PhysRevB.90.205422}. Since the number of non-zero unrelated components in the optical response tensors increases as the symmetry decreases, interest has grown in materials with intrinsically or extrinsically broken symmetries, such as the broken vertical symmetry in buckled monolayers \cite{siahin_monolayer_2009,PhysRevB.99.085432,le_fracture_2021,kezerashvili_effects_2021,PhysRevB.107.235416}, Janus materials \cite{Lu2017,Zhang2017,doi:10.1021/acs.nanolett.0c03412,doi.org/10.1002/lpor.202100726,Shi2023}, biased homobilayers \cite{yu_charge-induced_2015,brun_intense_2015,klein_electric-field_2017,zheng_gate_2023,PhysRevB.110.085433} and heterobilayers \cite{Gong2014,doi:10.1126/science.aac7820}, or the broken in-plane symmetries created by lattice strain \cite{Peng2020,Du2021,PhysRevB.106.125303} or an intrinsic lattice anisotropy, such as the one present in phosphorene \cite{Castellanos-Gomez_2014,PhysRevB.89.245407,Cho2017,PhysRevB.102.115305}. 
	
	In the past few years, since the isolation of the semiconducting 2D magnets \ce{CrI3} \cite{Huang2017} and \ce{Cr2Ge2Te6} \cite{Gong2017} in 2017, it has been an explosion in the discovery of new 2D materials with magnetic properties \cite{shi_review_2024}. While \ce{CrI$_3$} \cite{Gibertini2019} and \ce{Cr2Ge2Te6}, and also \ce{CrBr$_3$} \cite{Zhang_CrBr3_2019}, present a ferromagnetic order out-of-plane, there are also materials that are in-plane ferromagnets such as \ce{CrCl3} \cite{Pinto_CrCl3_2021} and \ce{CrSBr} \cite{wilson2021interlayer,lee2021magnetic,klein2023}. In particular, \ce{CrSBr} has a great potential for spintronics applications in nanoscale devices \cite{Burch2018}, in which the use of the spin degrees of freedom permit the manipulation and communication of information with extremely low power consumption \cite{10.1002/adma.200900809,Nikitov:2015,Chumak2015,10.1021/acs.nanolett.6b03052,Huang2017,kajale_twodimensional_2024}. Quantized collective spin excitations, known as magnons, are ideal quasi-particles for encoding spin information and developing spin-wave-based quantum computing concepts \cite{Smejkal2018,Rodin2020}. 
	
	\ce{CrSBr} is a semiconductor whose direct bandgap has been reported as $E_g\approx1.5\,\mathrm{eV}$ and an A-type antiferromagnet (AFM), with a magnetic easy axis along the $B$-direction \cite{klein2023,katscher1966chalkogenidhalogenide,goser1990magnetic,wang2019family,wang2020electrically,telford2020layered,lee2021magnetic,wilson2021interlayer,yang2021triaxial,klein2022control,klein2022sensing,telford2022coupling,wu2022quasi,lopezpaz_dynamic_2022,torres2023probing,qian_anisotropic_2023}, see Figure \ref{fig:CrSBr_unitcell}.
	Significantly, \ce{CrSBr} is a relatively stable material in air \cite{torres2023probing} for which recent works have shown structural phase transformations \cite{klein2022control}, correlated magneto-optical \cite{wilson2021interlayer,klein2022sensing,torres2023probing} and magneto-transport \cite{telford2020layered,telford2022coupling,wu2022quasi} properties, and highly anisotropic electron conductivity \cite{wu2022quasi}. 
	
	Two-dimensional materials offer an excellent platform for tuning their properties through strain, which has a pronounced impact on magnonic behavior \cite{wang_strain_2016,sadovnikov_magnon_2018,Esteras2022,ren_strain_2023}. Strain also significantly influences the optical response of these materials, particularly the characteristics of excitons \cite{quintela2025}. Recent research has uncovered strong interactions between excitons and magnons quasiparticles in 2D magnetic systems \cite{onga_antiferromagnetsemiconductor_2020,bae2022exciton,diederich_tunable_2023,sun2024dipolar,diederich_exciton_2025,datta_magnon_2025,belvin_exciton_2021,wang_exciton-magnon_2023,varela_ultrafast_2025}, an effect absent in conventional three-dimensional magnets. This effect opens new avenues for controlling the magnetic response of 2D semiconducting magnets through light-induced excitations. These findings highlight the need for advanced theoretical frameworks to accurately describe strain effects on excitons in 2D magnetic materials. State-of-the-art theoretical approaches typically rely on the Bethe-Salpeter Equation (BSE) to describe excitons and magnons \cite{marsili_spinorial_2021,esquembre_magnons_2025}. Such methods have been successfully applied to study excitons and optical phenomena in \ce{CrSBr}, including magneto-optical effects and linear dichroism \cite{klein2023,qian_anisotropic_2023}. However, strain effects on \ce{CrSBr} excitons were not investigated.

	In this work, we investigate the impact of strain on the linear optical response of \ce{CrSBr}. This study represents a first step toward exploring exciton-magnon coupling and understanding strain-induced effects relevant for spintronic applications. We begin in Section \ref{sec:lattice} by defining the lattice structure of \ce{CrSBr}, discussing the overall shape of its band structure and providing a brief overview on the effects of lattice strain in this material. In Section \ref{sec:bse}, we outline the methodology that will be considered to obtain the excitonic states, namely the BSE, together with the necessary modifications to the electrostatic interaction such that the intrinsic anisotropy of the lattice can be properly accounted for. Finally, in Section IV, we analyze the effects of strain on the diagonal components of the linear conductivity, together with the response for circularly polarized light. 
	
	
	\section{\label{sec:lattice}Lattice structure}
	Monolayer \ce{CrSBr} is a layered van der Waals antiferromagnet with a rectangular lattice  structure \cite{Esteras2022,klein2023}. It essentially consists of two buckled rectangular planes of \ce{CrS} fused together, with both surfaces capped by \ce{Br} atoms. Each atomic site sits at a linear combination of the Bravais vectors
	\begin{figure}
		\centering
		{
			\includegraphics[width=0.6\columnwidth]{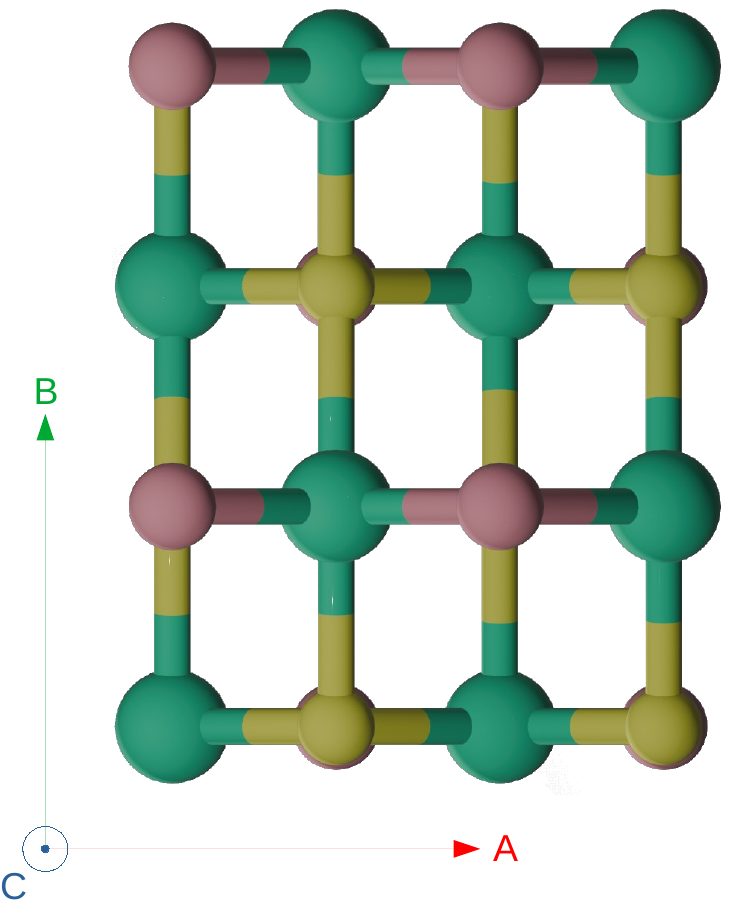}
		}
		\begin{minipage}[b]{0.35\columnwidth}
			\centering
			{
				\includegraphics[clip, width=\columnwidth]{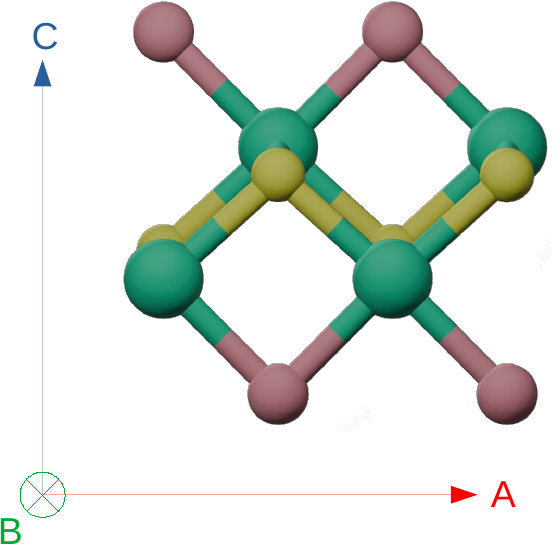}
			} \hfill
			{
				\includegraphics[width=\columnwidth]{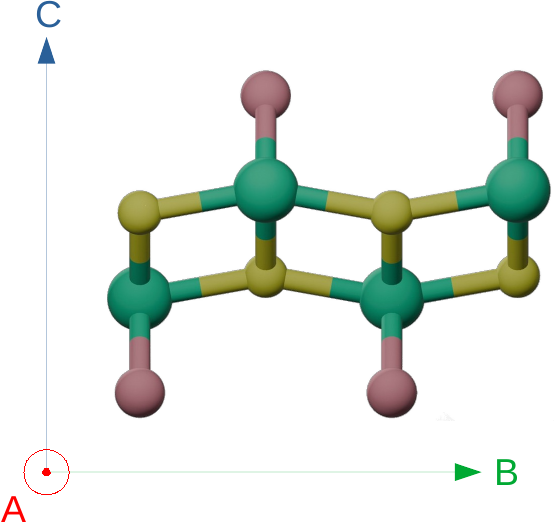}
			}
		\end{minipage}
		\caption{Top (left) and side views (right) of the unit-cell of a single $\mathrm{CrSBr}$ layer, where green, yellow, and pink represent the \ce{Cr}, \ce{S}, and \ce{Br} atoms, respectively. In the plots, the $A$, $B$, and $C$ directions correspond to $x$, $y$, and $z$ axis, respectively.}
		\label{fig:CrSBr_unitcell}
	\end{figure}
	\begin{equation}
		\textbf{a}_1 = (a,0,0), \ \ \ \textbf{a}_2 = (0,b,0),\label{eq:lattice_vectors}
	\end{equation}
	where $ a \approx 3.59\,\text{\AA}$ and $ b \approx 4.81\,\text{\AA}$. As seen in Fig. \ref{fig:CrSBr_unitcell}, each unit-cell counts with 3 pairs of \ce{Cr}, \ce{S}, and \ce{Br} located at
	\begin{align}
		\bm{\gamma}_1 & = \textbf{0}, \ \ \ \ \ \ \ \ \ \ \ \ \ \ \ \ \bm{\gamma}_2 = \frac{\textbf{a}_1 + \textbf{a}_2}{2} + h \hat{\textbf{z}} \ \ \ \text{(Cr)}\nonumber
		\\
		\bm{\sigma}_1 & = \frac{\textbf{a}_1}{2} + d_{\text{Cr-S}} \hat{\textbf{z}}, \ \ \bm{\sigma}_2 = \bm{\gamma}_2 - \bm{\sigma}_1 \ \ \ \ \ \ \ \ \ \ \ \text{(S)}
		\label{eq:S_sublattice_vectors}
		\\
		\bm{\beta}_1 & = \frac{\textbf{a}_1}{2} - d_{\text{Cr-Br}} \hat{\textbf{z}}, \ \bm{\beta}_2 = \bm{\gamma}_2 - \bm{\beta}_1 \ \ \ \ \ \ \ \ \ \ \text{(Br)},\nonumber
	\end{align}
	where $ h \approx 2.08\,\text{\AA}$ is the difference in height between the two Chromium atoms, $ d_{\text{Cr-S}} \approx 1.61\,\text{\AA}$, and $ d_{\text{Cr-Br}} \approx 1.77\,\text{\AA}$.
	
	\begin{figure}
		\centering\includegraphics[scale=0.385]{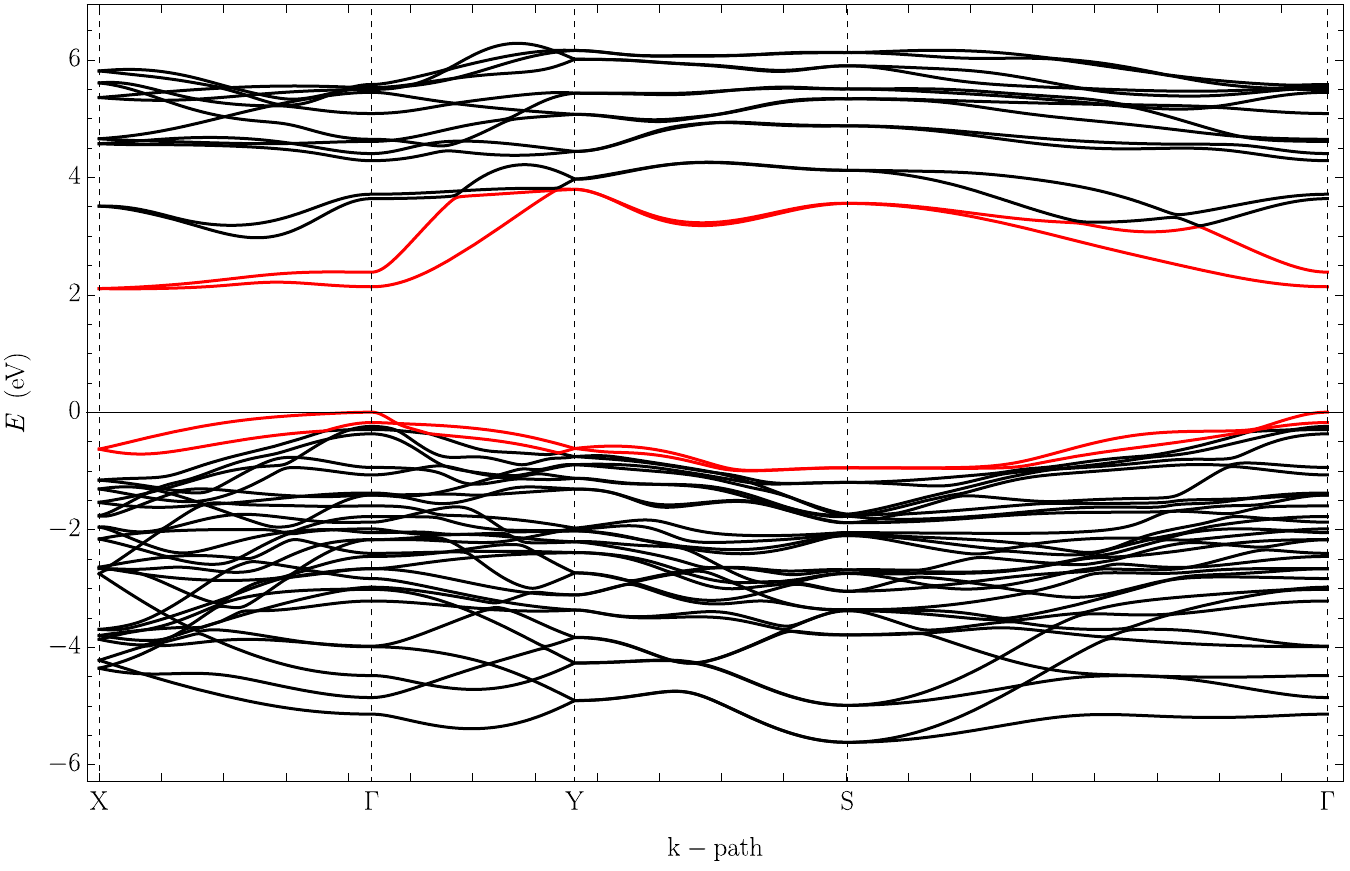}\,
		\includegraphics[scale=0.29]{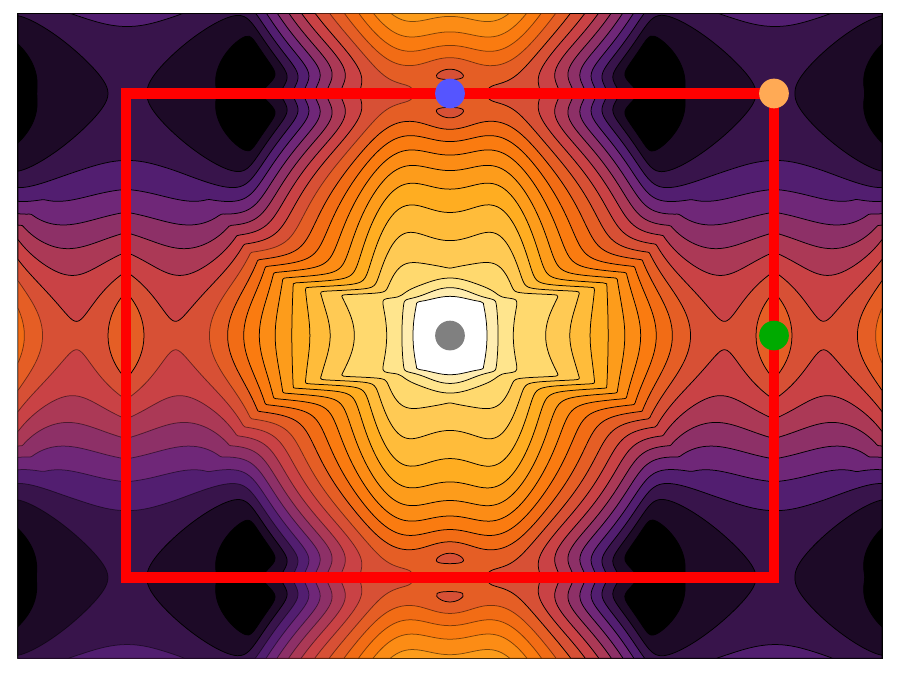}\includegraphics[scale=0.29]{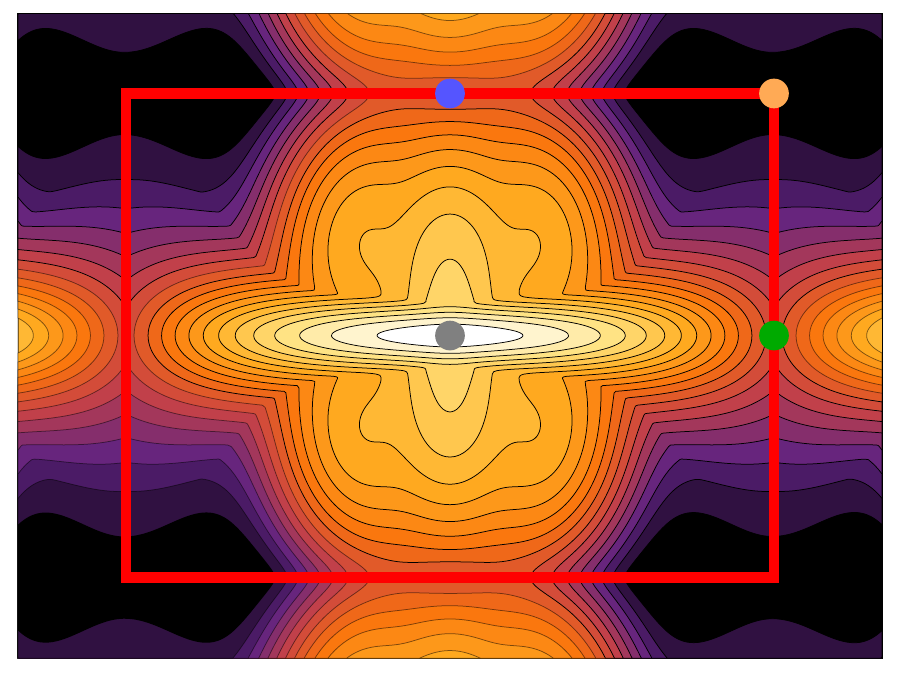}\,
		\includegraphics[scale=0.29]{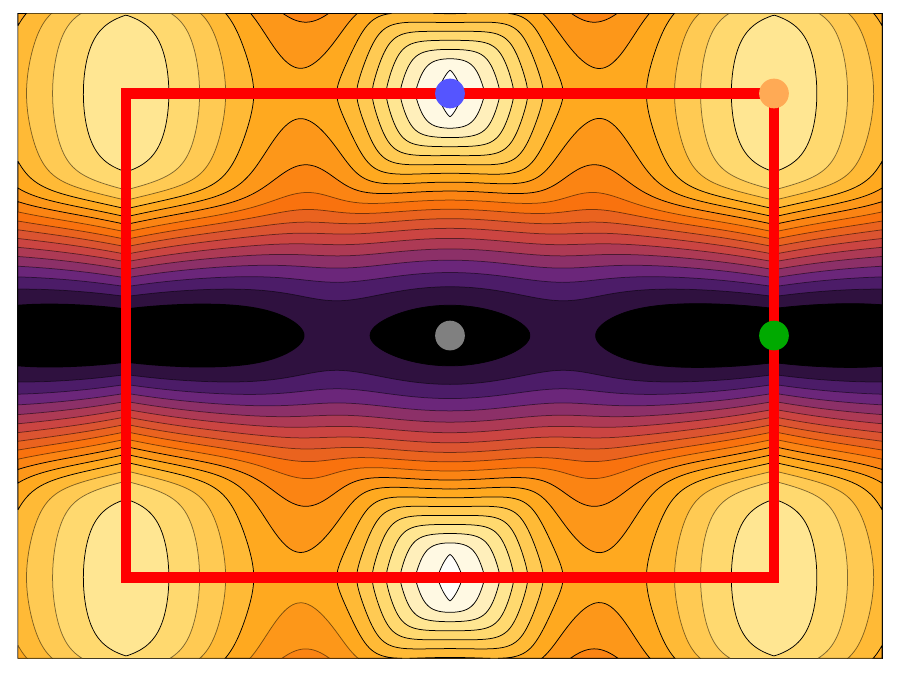}\includegraphics[scale=0.29]{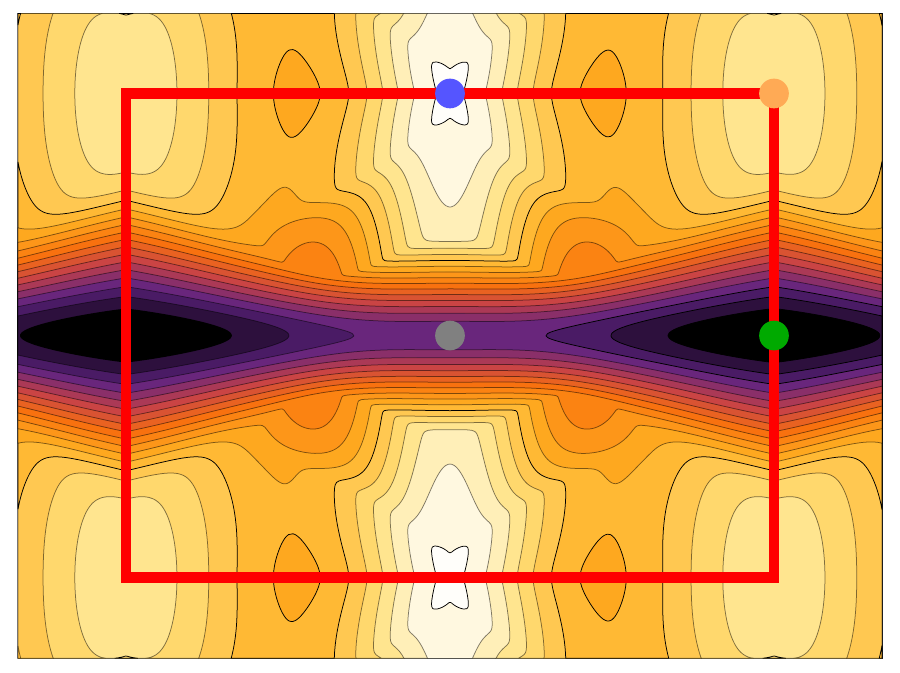}
		\vspace{-0.5cm}\centering\caption{Top: Band structure of \ce{CrSBr} along the $\mathrm{X}-\Gamma-\mathrm{Y}-\mathrm{S}-\Gamma$ path in the first Brillouin zone. Red bands were directly utilized for the BSE calculations in this work. Bottom: Contour plots of highlighted  bands (increasing in energy from top left to bottom right) in the FBZ (red rectangle). Gray, green, blue, and orange points represent the high--symmetry $\Gamma$, $\mathrm{X}$, $\mathrm{Y}$, and $\mathrm{S}$ points, respectively. \label{fig:bands}}
	\end{figure}
	
	The high-symmetry points of the rectangular lattice have coordinates: 
	\begin{align}
		\boldsymbol{\Gamma}&=[0,0] &\mathbf{X} & =[1/2,0]\nonumber\\
		\mathbf{Y}&=[0,1/2] &\mathbf{S} & =[1/2,1/2]\label{eq:high-sym}
	\end{align}
	in the basis of the reciprocal lattice vectors  $\mathbf{b}_1$ and $\mathbf{b}_2$. These points are represented in Fig. \ref{fig:bands} by the gray, green, blue, and orange points, respectively. 
	
	For our calculations, we consider a tight-binding-like Hamiltonian written in a basis of maximally localized Wannier functions (MLWF) \cite{MOSTOFI20142309,Pizzi_2020}, as obtained and discussed in Ref. \cite{Esteras2022}, where the preferential magnetization direction was set along the $y$-axis. The wannerization was performed on DFT+U calculations, where U is the on-site Coulomb repulsion. The exchange-correlation energy was described by the generalized gradient approximation (GGA) and the Perdew-Burke-Ernzerhof (PBE) functional. The band structure of this Hamiltonian is portrayed in Fig. \ref{fig:bands}, where a scissor cut has been applied to match the direct bandgap with other theoretical works at $E_g=2.1\,\mathrm{eV}$\cite{qian_anisotropic_2023,klein2023}. The top panel shows the energy following a path in $k$-space along the high-symmetry points of the material, see Eq. (\ref{eq:high-sym}). The bottom panel portrays the contour plots of the two highest valence bands and two lowest conduction bands in the first Brillouin zone (FBZ).

	
	
	\subsection{Strain}
	\label{sec:strain}
	
	\begin{figure*}
		\centering
		\includegraphics[scale=0.29]{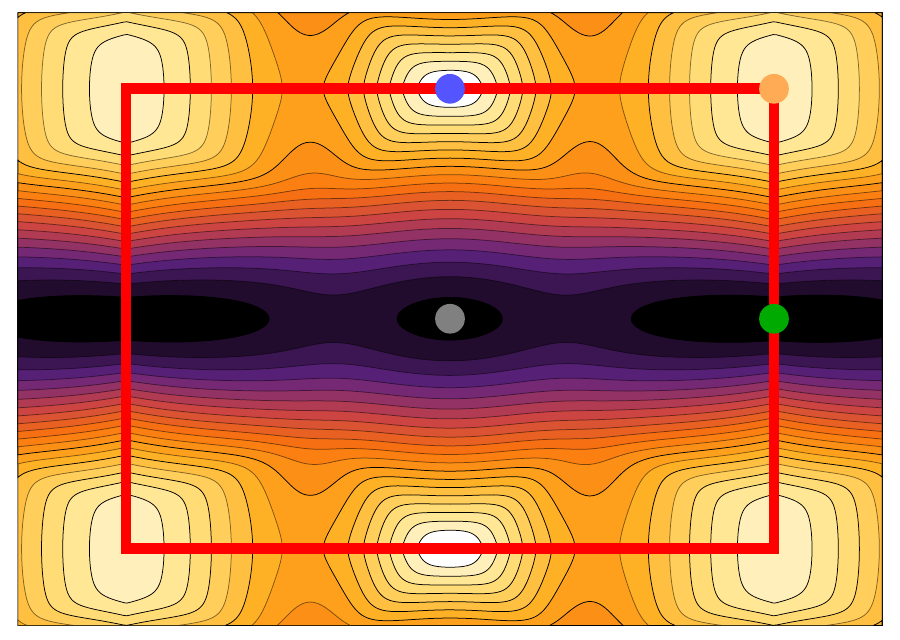}\,\includegraphics[scale=0.29]{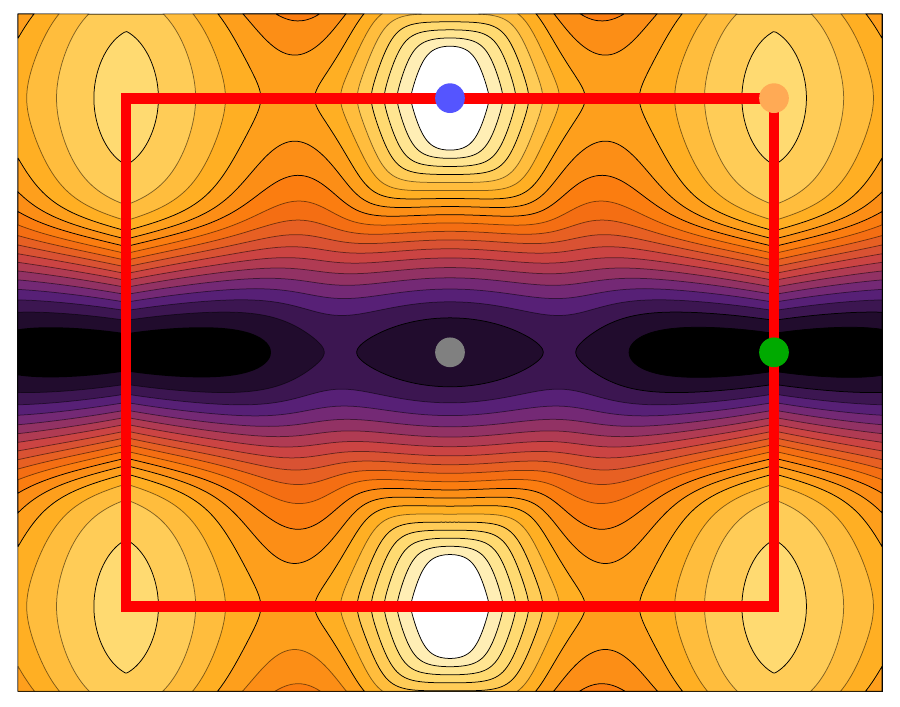}\,\includegraphics[scale=0.29]{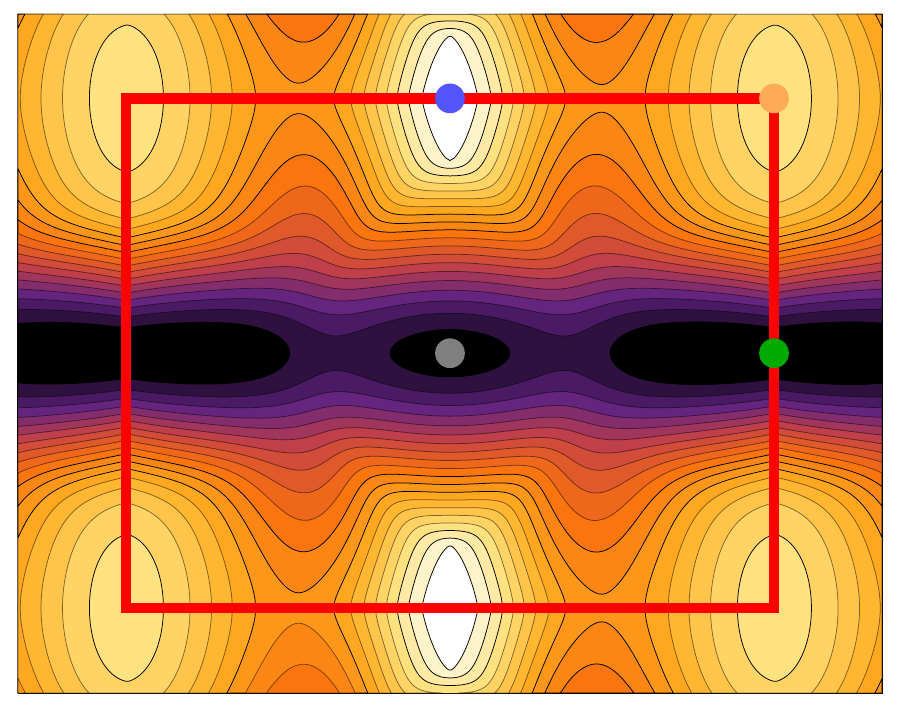}\,\includegraphics[scale=0.29]{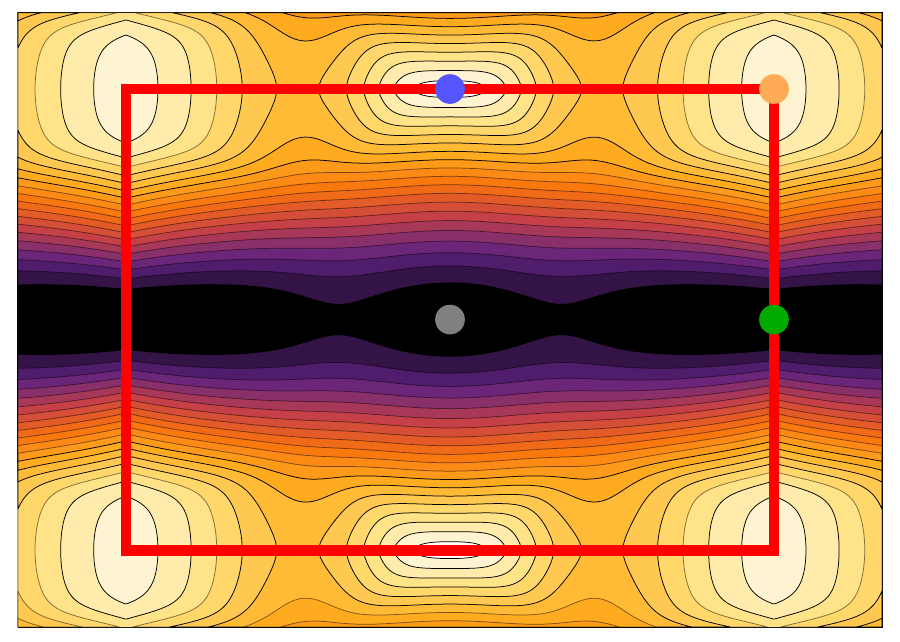}
		\includegraphics[scale=0.29]{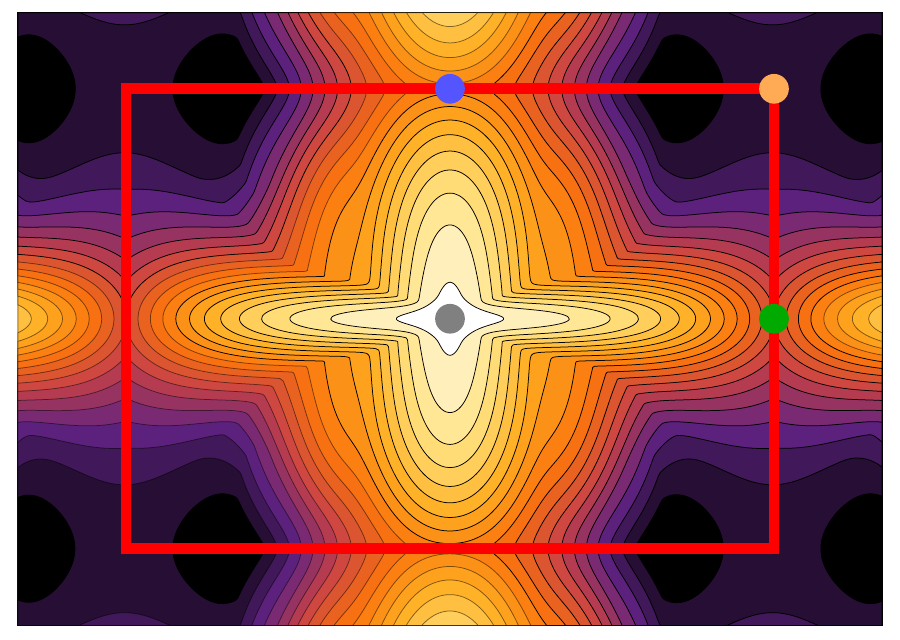}\,\includegraphics[scale=0.29]{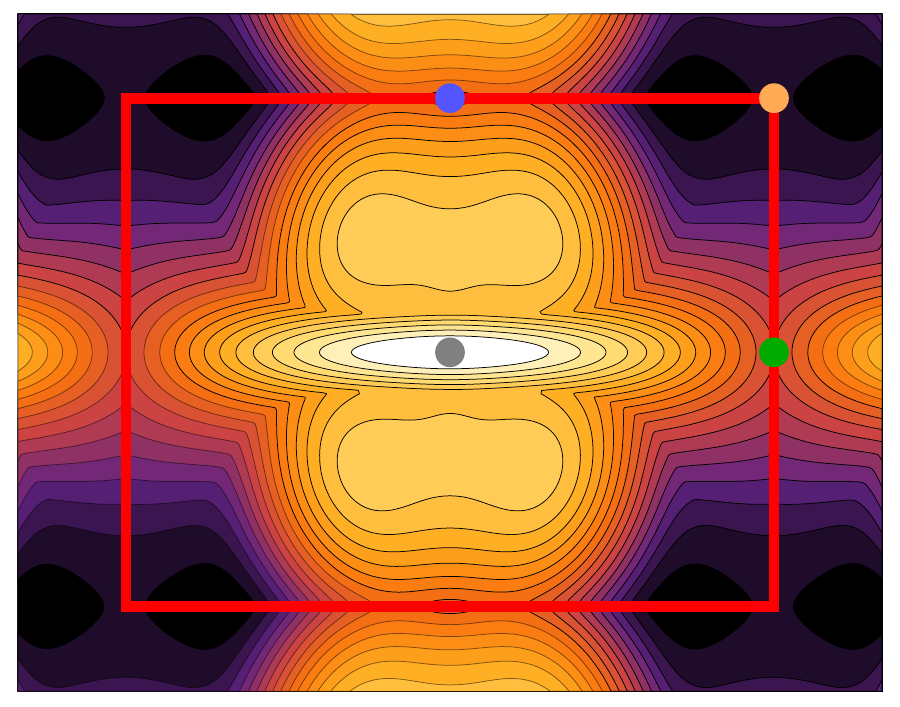}\,\includegraphics[scale=0.29]{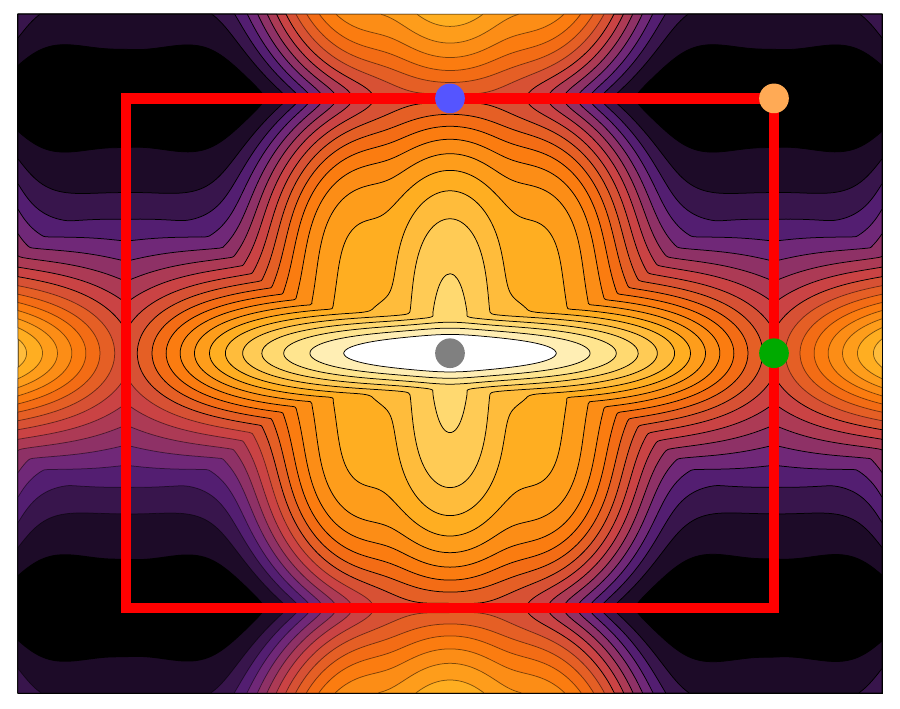}\,\includegraphics[scale=0.29]{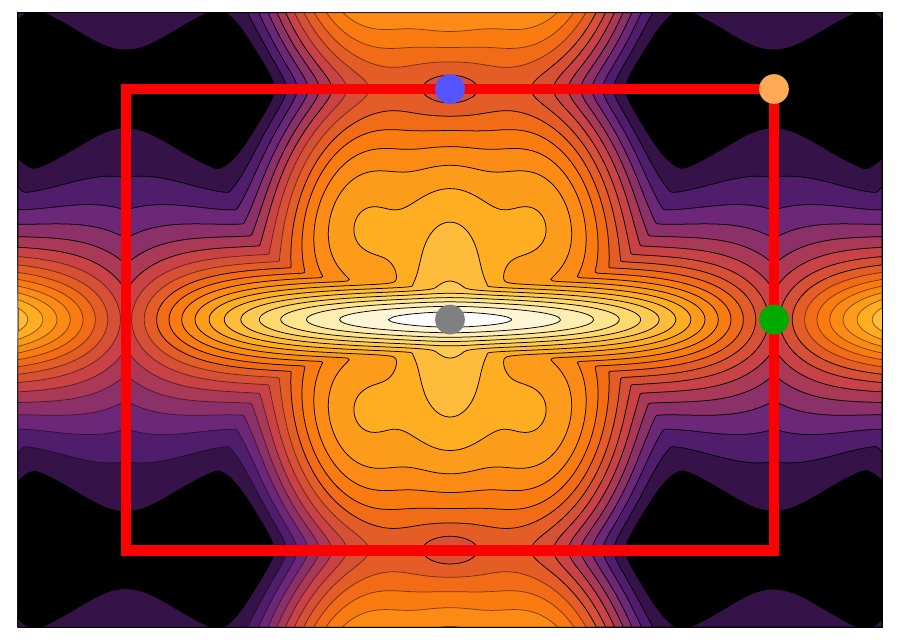}
		\vspace{-0.5cm}\centering\caption{Contour plots of the two bands closest to the Fermi level (conduction top, valence bottom) in the FBZ (red rectangle) for strain configurations $95\%\, A$, $105\%\, A$, $95\%\, B$, $105\%\, B$ (left to right respectively). \label{fig:bands_strain}}
	\end{figure*}
	
	The strain configuration will be described by its effects on the $\mathbf{a}_{1/2}$ vectors. Recalling the definition of the primitive vectors given by Eq. (\ref{eq:lattice_vectors}), we will identify the strain as $A$ or $B$ strain as it will directly affect the $a$ and $b$ lengths in Eq. (\ref{eq:lattice_vectors}). Therefore, besides the unstrained system, we will also consider compression ($95\%$ strain) and extension ($105\%$ strain) of the material along both $A$ and $B$ directions. The effects of strain on the first valence and conduction bands are presented in Fig. \ref{fig:bands_strain}. Note that the lowest energy region of the conduction band, which is like a cylindrical-shape valley along the $\Gamma$X direction, is modified by the strain configuration. As we will see in the following sections, this will have clear implications on exciton wavefunctions and, consequently, in the optical response of the material. More pronounced is the effect of strain on the valence band. The highest energy region is like a cylindrical-shape mountain along the $\Gamma$X direction, and the compression along the A direction clearly affects the energy between the $\Gamma$Y direction.

	
	
	\section{\label{sec:bse}Bethe-Salpeter equation}
	
	The excitation of a valence band electron onto the conduction band of a material, leaving behind a hole in the valence band, can lead to the formation of bound states. Those states are bounded via the Coulomb interaction of the electron-hole pair and their energy lie within the band-gap of the system. Using valence ($v$) and conduction ($c$) single-particle states, we expand exciton states of zero center of mass momentum ($\mathbf{Q} = 0$) in a basis of free electron-hole pairs as 
	\begin{equation}
		\left|X_N\right>=\sum_{vc\mathbf{k}}A_{vc\mathbf{k}}^{\left(N\right)}c^{\dagger}_{c\mathbf{k}}c_{v\mathbf{k}}\left|0\right>,
	\end{equation}
	where $\left|0\right>$ is the ground state, identified as the Fermi sea, $A_{vc\mathbf{k}}^{\left(N\right)}$ are the coefficients in the expansion of the excitonic states in terms of electron--hole excitations, and $N$ is a generic quantum number which describes the excitonic state. Following from Esteve-Paredes \emph{et al.} \cite{Esteve-Paredes2025}, exciton states are found by solving the BSE \cite{PhysRevB.62.4927} in the Tamm-Dancoff approximation (TDA) as
	\begin{align}
		E^{\left(N\right)}\,A_{vc\mathbf{k}}^{\left(N\right)} &=\left(E_{c\mathbf{k}}-E_{v\mathbf{k}}\right)A_{vc\mathbf{k}}^{\left(N\right)}+\nonumber\\
		&\quad+\sum_{v^{\prime}c^{\prime}\mathbf{k}^\prime}\left<vc\mathbf{k}\middle|K_\mathrm{eh}\middle|v^{\prime}c^{\prime}\mathbf{k}^\prime\right>A_{v^{\prime}c^{\prime}\mathbf{k}^\prime}^{\left(N\right)},
	\end{align}
	where $E_{c/v\mathbf{k}}$ are the conduction/valence electronic bands, and $K_{\mathrm{eh}}$ is the electron--hole interaction kernel. Following the numerical implementation of the BSE in the \textsc{XATU} code \cite{URIAALVAREZ2024109001}, the interaction kernel reads $K_{\mathrm{eh}}=-\left(D-X\right)$, where $D$ is the direct term, given by 
	\begin{eqnarray}
		D_{vcv^\prime c^\prime}\left(\mathbf{k},\mathbf{k}^\prime\right) = \hspace{5cm} \nonumber \\
		\int\psi_{c\mathbf{k}}^*\left(\mathbf{r}\right)\psi_{v^\prime\mathbf{k}^\prime}^*\left(\mathbf{r}^\prime\right)W\left(\mathbf{r},\mathbf{r}^\prime\right)\psi_{c^\prime\mathbf{k}^\prime}\left(\mathbf{r}\right)\psi_{v\mathbf{k}}\left(\mathbf{r}^\prime\right),\label{eq:direct}
	\end{eqnarray}
	and $X$ is the exchange term (obtained by interchanging $c^{\prime}\mathbf{k}^\prime$ and $v\mathbf{k}$). The $W\left(\mathbf{r},\mathbf{r}^\prime\right)$ represents the screened Coulomb interaction between two charged particles, which will be treated in the static limit as we detail in the following section.
	
	\subsection{Screened Coulomb interaction}
	\label{sec:screen}
	
	When solving the BSE for excitons in 2D materials \cite{,https://doi.org/10.1002/pssb.202200097,PhysRevB.105.045411,URIAALVAREZ2024109001,PhysRevB.110.085433,Esteve-Paredes2025}, the electrostatic interaction modeled by the Rytova-Keldysh (RK) potential \cite{rytova_screened_1967,keldysh_coulomb_1979} may represent a good description. The RK potential is derived from the Poisson's equation, in which the charge from the induced polarization in response to an external electric field is only contained in a 2D plane. In real space, this potential is given by
	\begin{equation}
		W_{\mathrm{RK}}\left(\mathbf{r},\mathbf{r}^\prime\right)=-\frac{\hbar c\alpha}{\epsilon_{r}}\frac{\pi}{2r_{0}}\left[H_{0}\left(\frac{\left|\mathbf{r}-\mathbf{r}^\prime\right|}{r_{0}}\right)-Y_{0}\left(\frac{\left|\mathbf{r}-\mathbf{r}^\prime\right|}{r_{0}}\right)\right],\label{eq:rytova-keldysh}
	\end{equation}
	where $H_{0}(x)$ is the zeroth-order Struve function, $Y_{0}(x)$ is the zeroth-order Bessel function of the second kind, $r_0$ is a material-specific screening length \cite{PhysRevB.84.085406}, and $\epsilon_{r}$ is the average dielectric constant of the medium surrounding the monolayer.  While most frequently studied materials, such as hBN \cite{Henriques_2020}, TMDCs \cite{PhysRevB.93.235435,PhysRevB.84.085406,Huang2016}, and biased bilayer systems \cite{Castro_2010,PhysRevB.105.115421} are well-modeled by an isotropic screening length, the high intrinsic anisotropy of $\mathrm{CrSBr}$ (see Fig. \ref{fig:CrSBr_unitcell}, for instance) demands we consider a more general form of screening. From the large anisotropy of the bands, as it can be observed in Fig. \ref{fig:bands}, we will consider different screening lengths along the $x$ and $y$ directions.

	Explicitly, the anisotropic Rytova--Keldysh potential will read as 
	\begin{equation}
		W_{\mathrm{RK}}\left(\mathbf{r},\mathbf{r}^\prime\right)=-\frac{\hbar c\alpha}{\epsilon_{r}}\frac{\pi}{r_{0}^x+r_{0}^y}\left[H_{0}\left(\rho\right)-Y_{0}\left(\rho\right)\right],\label{eq:aniso-rytova-keldysh}
	\end{equation}
	where the $r_0$ term usually outside of the Bessel and Struve functions is now taken as the average of the screening lengths along the $x$ ($r_{0}^x$) and the $y$ ($r_{0}^y$) direction. Here, 
	\begin{equation}
		\rho=\sqrt{\left(\frac{x-x^\prime}{r_{0}^x}\right)^2+\left(\frac{y-y^\prime}{r_{0}^y}\right)^2+\left(\frac{z-z^\prime}{r_{0}^z}\right)^2},
	\end{equation}
	where the $z$ component is explicitly included due to the complex lattice structure and $r_{0}^z$ is also taken as the average of the $x$ and $y$ components. While the material is still an atomically thin layer and, therefore, finite in the $z$ direction, it is important to define a ``vertical screening length'' as we will have distinct $z,z'$ positions for many of the atomic sites. Regarding the values of $r_{0}^x$ and $r_{0}^y$, these can be obtained by doing a linear fit of the dielectric function of the material (as obtained via DFT calculations) near $q=0$. Considering previous calculations of the dielectric function for monolayer CrSBr \cite{qian_anisotropic_2023}, a linear fit near $q=0$ along both $q_x$ and $q_y$ directions gives
	\begin{align}
		r_{0}^x &= 26.3894\,\text{\AA} &r_{0}^y &= 65.3451\,\text{\AA}.
	\end{align}
	Note the significant difference of the screen parameter depending on the direction. This presents an ``anisotropy ratio'' of $r_{0}^y/r_{0}^x \approx 2.476$, in good agreement with other computed results presented in the literature \cite{klein2023}.
	
	\begin{figure*}
		\centering
		\includegraphics[scale=0.64]{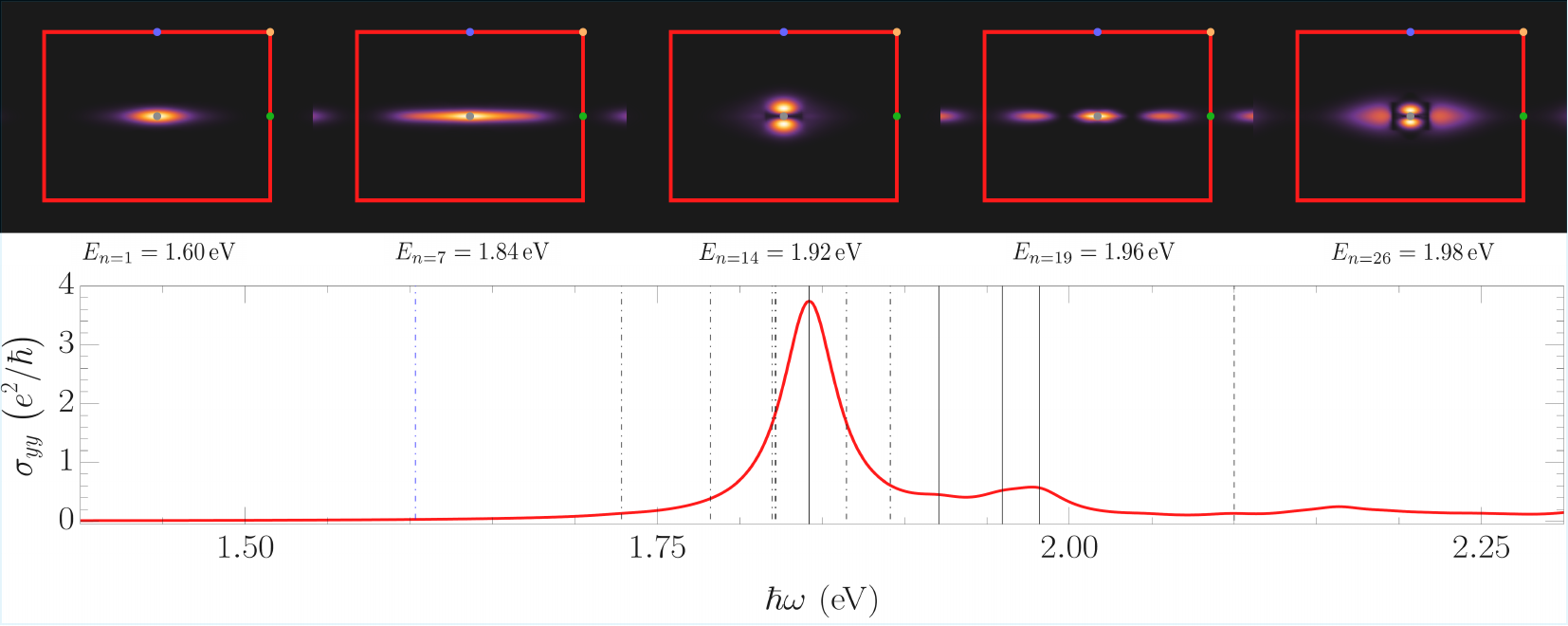}
		\centering\caption{Wave functions of the $n=1$ (optically dark), $n=7$, $n=14$, $n=19$, and $n=26$ exciton states plotted in the FBZ (top panel), together with the subgap region of the optical conductivity (bottom panel). In the bottom panel, vertical dashed line at $\hbar\omega=2.1\,\mathrm{eV}$ corresponds to the bandgap at the $\Gamma$ point for the unstrained system, solid vertical lines correspond to the optically bright states and dot-dashed lines correspond to optically dark states until $n=7$ (blue dot-dashed line highlights the lowest energy exciton state, $n=1$).  \label{fig:wave-functions}}
	\end{figure*}
	
	Knowing the specific form of the anisotropic Rytova--Keldysh potential, we can now proceed to compute the excitonic wave functions. While the large anisotropy means angular momentum is not a good quantum number in this system, we can still label the excitonic states by their energy ordering. In Fig. \ref{fig:wave-functions}, we plot the exciton wave functions for the lowest energy state ($n=1$), which is optically dark, together with the first four optically bright states ($n=7,\,14,\,19,\,26$). Although it will be discussed in the following section, the states are presented with a snippet of the sub-gap optical conductivity for easier visualization, with solid and dot-dashed lines highlighting the spectral location of the optically bright and dark states, respectively. We only represent the first eight optically dark states ($n=1$ state highlighted in blue for easier identification). Note that 7-exciton and 26-exciton strongly contribute to the optical conductivity, and while 7-exciton is quite spread in the k$_y$ direction, i.e. it behaves as a 1D exciton, the 26-exciton starts to have some spread along the k$_x$ direction. 
	
	While there is a resemblance of the found excitonic states with previous calculations in the literature \cite{qian_anisotropic_2023}, there are important differences, particularly in the energies. This can be easily understood by analyzing the used bands structure, which significantly differ. As a consequence, the resulting Wannier Hamiltonian will be quite distinct and, therefore, the excitonic states resulting from solving the BSE in the two systems are difficult to compare.
	

	
	\section{\label{sec:optical} Strain Effects on the Linear Optical Response}
	
	\begin{figure*}
		\centering
		\includegraphics[scale=0.66]{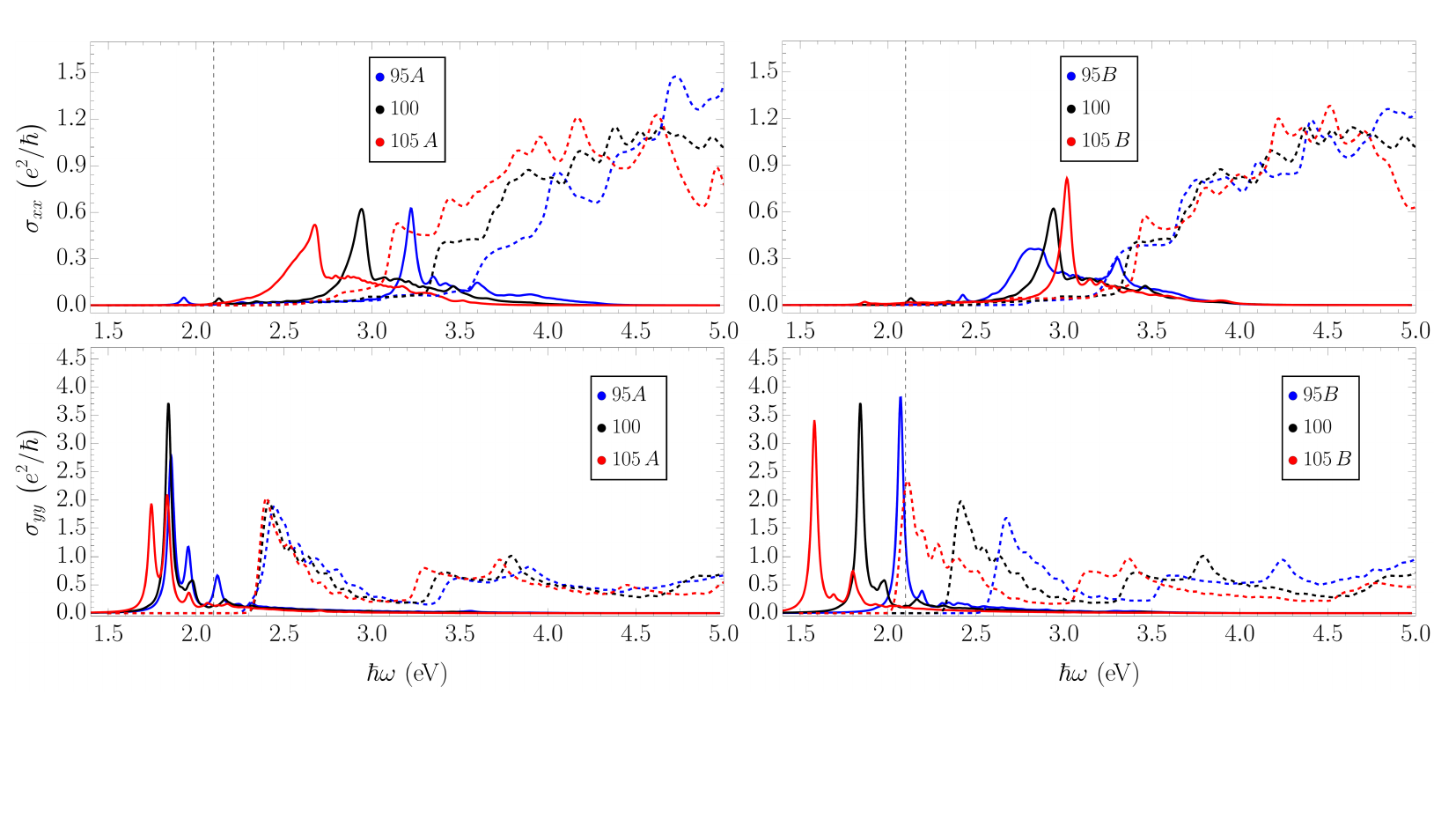}
		\vspace{-2cm}\centering\caption{Excitonic (solid lines) and IPA (dashed lines) contribution to the diagonal tensor elements of the linear optical conductivity (top: $\sigma_{xx}$, bottom: $\sigma_{yy}$). Left panels correspond to strain in the $A$ direction ($x$-axis), while right panels correspond to strain in the $B$ direction ($y$-axis). Blue, black, and red lines correspond to $95\%$ strain, unstrained, and $105\%$ strain in the corresponding column's direction, respectively. In all panels, vertical dashed line at $\hbar\omega=2.1\,\mathrm{eV}$ corresponds to the bandgap at the $\Gamma$ point for the unstrained system. \label{fig:conductivity}}
	\end{figure*}
	
	Having examined both the Bethe-Salpeter equation and the role of anisotropy in the electrostatic electron-hole interaction, we now turn to the influence of strain on the linear optical response of $\mathrm{CrSBr}$.
	
	
	To understand the impact of strain on the linear optical response of the 2D magnetic material, it is instructive to examine the optical response in the absence of electron-hole interactions, i.e. within the independent particle approximation (IPA). The IPA calculations allow us to isolate the primary changes in optical conductivity arising from modifications in the energy band dispersion. This, in turn, provides an initial information to further analyze and disentangle the excitonic effects under strain. Accordingly, we compute the linear optical response at both the IPA level and with full excitonic interactions.
	
	The excitonic optical conductivity is computed directly from the eigenstates of the Bethe-Salpeter equation as
	\begin{eqnarray}
		\frac{\sigma_{\alpha\beta}\left(\omega\right)}{\sigma_{0}}=\frac{-i}{2\pi^{3}}\sum_{n}\left[\frac{E_{n}X_{0n}^{\alpha}X_{n0}^{\beta}}{E_{n}-\hbar\omega}-\left(\omega\rightarrow-\omega\right)^{*}\right],
		\label{eq:conductivity}
	\end{eqnarray}
	where $\sigma_{0}=\frac{e^2}{4\hbar}$, $\alpha/\beta$ are Cartesian directions, $E_{n}$ is the energy of the excitation, in which an imaginary broadening $\Gamma=20\,\mathrm{meV}$ is introduced via the transformation $E_{n} \rightarrow E_{n} - i\Gamma$ to model a realistic spectrum. The generic exciton dipole matrix element reads \cite{pedersen_intraband_2015,taghizadeh_nonlinear_2019,PhysRevB.107.235416}
	\begin{eqnarray}
		X_{0n}^{\alpha}=i\frac{\hbar}{m_{0}}\sum_{c,v}\int\,A_{vc\mathbf{k}}^{\left(n\right)}\frac{p_{vc\mathbf{k}}^{\alpha}}{E_{c\mathbf{k}}-E_{v\mathbf{k}}}d^{2}\mathbf{k},\label{eq:optical_dipole}
	\end{eqnarray}
	with $A_{vc\mathbf{k}}^{\left(n\right)}$ the exciton wavefunction between conduction band $c$ and valence band $v$, $p_{vc\mathbf{k}}^{\alpha}$ the interband momentum matrix element between the same band pair, defined as 
	\begin{align}
		p_{vc\mathbf{k}}^{\alpha} = &\frac{m_{0}}{\hbar}\left<v,\mathbf{k}\middle|\frac{\partial\mathcal{H}\left(\mathbf{k}\right)}{\partial k_{\alpha}}\middle|c,\mathbf{k}\right>\nonumber
		\label{eq:p_circ},
	\end{align}
	and $E_{c/v\mathbf{k}}$ the dispersion of the bands in question. 
	
	The IPA contribution to the linear optical response will be obtained by directly applying the Kubo formalism as defined in Refs. \cite{pedersen_intraband_2015,URIAALVAREZ2024109001} and computed with the \textsc{Opticx} code \cite{Esteve-Paredes2025}\footnote{The \textsc{Opticx} package can be found within the repository of the \textsc{XATU} code at \texttt{https://github.com/xatu-code}.}.
	
	As the dominant contribution to the sub-gap response will stem from the bands closest to the Fermi level, we perform the excitonic computations while taking into account the two valence and two conduction bands highlighted in red in Fig. \ref{fig:bands}. This decreases the computational cost without compromising the accuracy of the calculations and it makes possible to consider a greater number of neighboring unit cells in the calculation. Considering $N=90$ neighboring cells and an imaginary broadening of $\Gamma=0.025\,\mathrm{eV}$, we compute the linear optical conductivity under both $A$ and $B$ strain, see Fig. \ref{fig:conductivity}.
	
	Figure \ref{fig:conductivity} portrays the $xx$ (top panels) and $yy$ (bottom panels) tensor components of the linear conductivity tensor for five distinct strain configurations: an unstrained system, $95\%$ strain along both $A$ and $B$ directions, and $105\%$ strain along the same directions. In this figure, the panels in  the diagonal (top-left and bottom-right) correspond to strains parallel to the tensor component of the linear conductivity, while the off-diagonal panels (top-right and bottom-left) correspond to strains perpendicular to the tensor component of the linear conductivity.

	First we note that the excitonic optical conductivity presents strong excitons peaks within the bandgap for $y$ polarization ($B$ axis), while excitonic peaks for $x$ polarization are very small. This is in agreement with what is observed in experiments, see for example Ref. \cite{datta_magnon_2025}. This results in significant birefringence or linear dichroism for photon energies within the energy gap.

	By first examining the IPA calculations, Fig. \ref{fig:conductivity} reveals that strain effects are significantly more pronounced when the strain direction is collinear with the diagonal component of the linear optical response under consideration. This is because strain alters the orbitals involved in bonding along that direction, thereby modifying the band energy structure. The primary manifestation of these changes is a redshift or blueshift in energy, corresponding to tensile or compressive strain, respectively, which are consistent with variations in the band gap. The excitonic optical conductivity also inherits the same behaviour, with its dominant feature being a redshift or blueshift depending on the sign of the applied strain.

	When examining the effects of strain applied perpendicular to the diagonal component of the linear optical response, we find that the IPA optical conductivity remains largely unchanged. However, significant modifications appear in the excitonic states, particularly those within the band gap along the $y$-direction. This behaviour can be understood by analyzing the exciton wavefunctions shown in Fig. \ref{fig:wave-functions}, especially bright excitons that contribute to the optical response of the system. These wavefunctions are localized in regions of $k$-space where the band energies are sensitive to strain, as illustrated in the conduction and valence energy bands in Fig. \ref{fig:bands_strain}. This is particularly visible under tensile strain, in which we observe a strong redshift energy. Although bound excitons in \ce{CrSBr} exhibit quasi-1D behaviour along the $y$ ($B$) direction, they are clearly influenced by strain applied along the $x$ ($A$) direction.

	Moreover, the lineshape of the excitonic optical conductivity evolves under strain, an effect attributed to the high sensitivity of exciton wavefunctions to lattice distortions. For instance, the wavefunctions of excitons $7$ and $26$ in the unstrained system differ notably, exciton $7$ is primarily extended along the $k_y$-direction, while exciton $26$ has a stronger contribution along the $k_x$-direction. This wavefunction distribution implies a distinct response to strain, with direct consequences for the energy positions of the excitonic series, see more details in Appendix \ref{sec:appendix2}, and the corresponding oscillator strengths.

	\begin{figure*}
		\centering\includegraphics[scale=0.65]{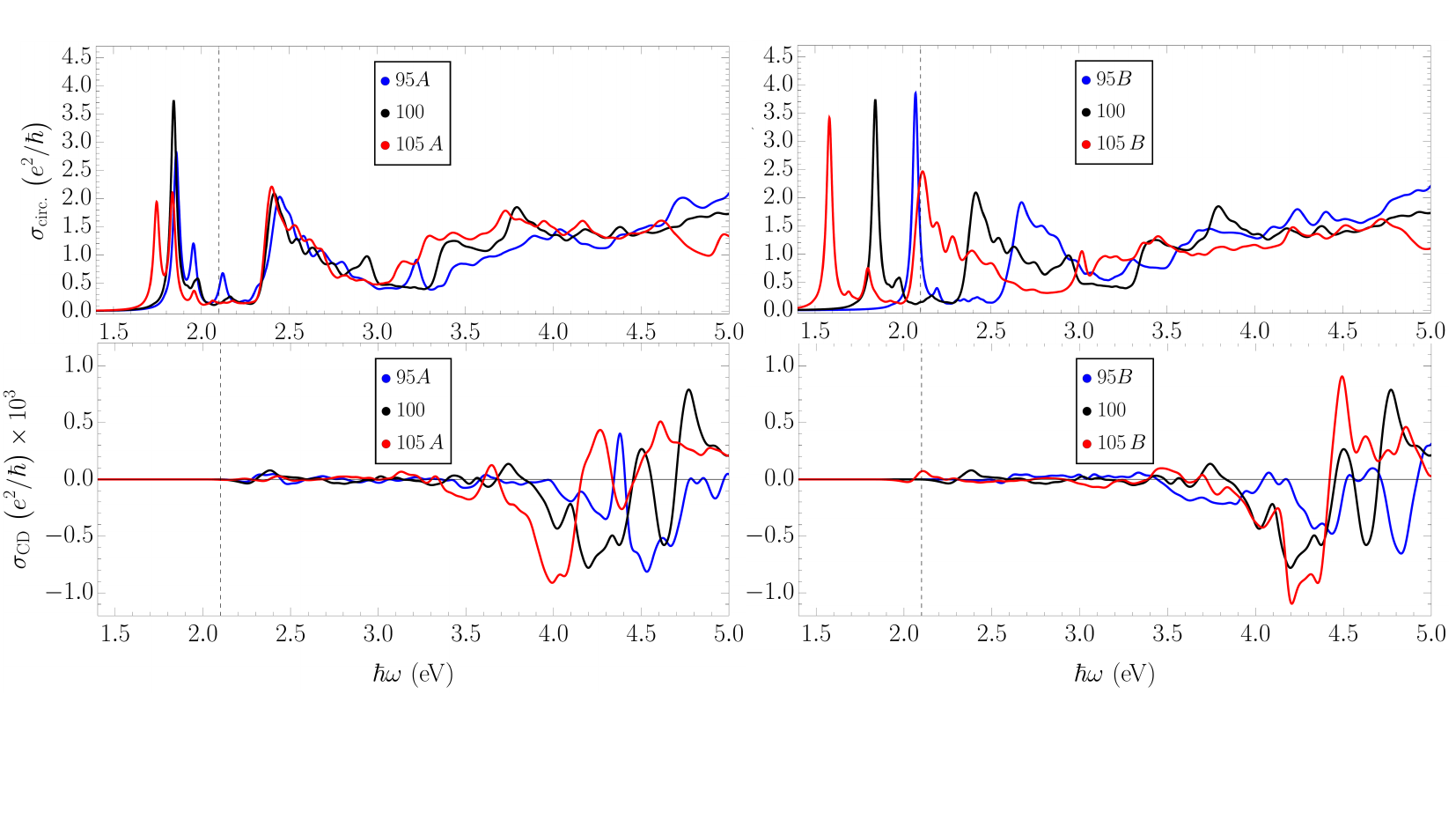}
		\vspace{-2cm}\centering\caption{Linear optical conductivity with both excitonic and IPA contributions for circularly polarized light. Left panels correspond to strain in the $A$ direction ($x$-axis), while right panels correspond to strain in the $B$ direction ($y$-axis). Top panels correspond to left-handed circular polarization, while bottom panels correspond to difference between left- and right-handed polarization. Blue, black, and red lines correspond to $95\%$ strain, unstrained, and $105\%$ strain in the corresponding column's direction, respectively. In both panels, vertical dashed line at $\hbar\omega=2.1\,\mathrm{eV}$ corresponds to the bandgap at the $\Gamma$ point for the unstrained system. \label{fig:circ_pol}}
	\end{figure*}

	Now we consider the optical response to circularly polarized light, which is related to the magnetic circular dichroism (MCD), see Fig. \ref{fig:circ_pol}. The circular optical conductivity can be related to the off-diagonal components of the linear optical conductivity. Simplifying Eq. (\ref{eq:optical_dipole}) for circularly polarized light, we have
	\begin{align}
		X_{0n}^{\pm}&=\frac{i}{\sqrt{2}}\sum_{c,v}\int\,A_{vc\mathbf{k}}^{\left(n\right)}\frac{\left<v,\mathbf{k}\middle|\left[\frac{\partial\mathcal{H}\left(\mathbf{k}\right)}{\partial k_{x}}\pm i\frac{\partial\mathcal{H}\left(\mathbf{k}\right)}{\partial k_{y}}\right]\middle|c,\mathbf{k}\right>}{E_{c\mathbf{k}}-E_{v\mathbf{k}}}d^{2}\mathbf{k}\nonumber\\
		&=\frac{i}{\sqrt{2}}\left[X_{0n}^{x}\pm iX_{0n}^{y}\right],
	\end{align}
	meaning that circular polarization will, in general, mix the four in-plane tensor components as
	\begin{eqnarray}
		\sigma_{\mathrm{\pm}}\left(\omega\right) & =\frac{\sigma_{xx}\left(\omega\right)}{2}+\frac{\sigma_{yy}\left(\omega\right)}{2} \pm \frac{i}{2}\left[\sigma_{yx}\left(\omega\right)-\sigma_{xy}\left(\omega\right)\right].
		\label{eq:circularconductivity}
	\end{eqnarray}
	When all four tensor components are mixed, the effects of strain become considerably more complex. However in monolayer \ce{CrSBr}, the $yy$ tensor component is the dominant one until $\hbar\omega \approx3-3.5\,\mathrm{eV}$ and, examining Eq. (\ref{eq:circularconductivity}), one expects the circular optical conductivity under strain to closely resemble the features observed in the $yy$ component of the linear optical conductivity within this frequency range. The MCD signal, which is computed by taking the difference of the left- and right-handed circular optical conductivity, is found to be very small within the band gap region, see bottom panels of Fig. \ref{fig:circ_pol}. The MCD is directly connected to the $xy$ and $yx$ tensor components, which are anti-symmetrical between themselves. The small MCD signal is understood because the $xy$ and $yx$ tensor components are $3$ orders of magnitude smaller than the $xx$ and $yy$ tensor components. Additionally, while a subtle difference is observable below the bandgap (more clearly visible when considering strain in the $B$ direction), the effect is much smaller than what is observed in the single-particle regime and, therefore, is barely observable even when magnified by a factor of $10^3$. As a final comment, the $xy$ and $yx$ tensor components are also responsible for magneto-optical phenomena such as the Faraday and Kerr effects. Due to their small magnitude, these effects are likewise weak, consistent with experimental observations.
	
	
	In the ground state, the magnetization orientation in \ce{CrSBr} lies along the $y$ (B) direction. This orientation, which can be modified by an external magnetic field, influences the optical and the magneto-optical response of the material. In Appendix \ref{sec:appendix1}, we compare the effects of the three distinct magnetization orientations on the the diagonal tensor components of the conductivity and the circular dichroism. Notably, we observe that the MCD signal becomes two orders of magnitude stronger above the band gap when the magnetization is aligned along the $z$ direction. This enhancement is consistent with observations in other materials, such as \ce{CrI3}, where out-of-plane magnetization leads to strong MCD signals \cite{wu_physical_2019}.
	
	
	\section{\label{sec:conclusion}Summary}

	\begin{figure*}
		\centering\includegraphics[scale=0.64]{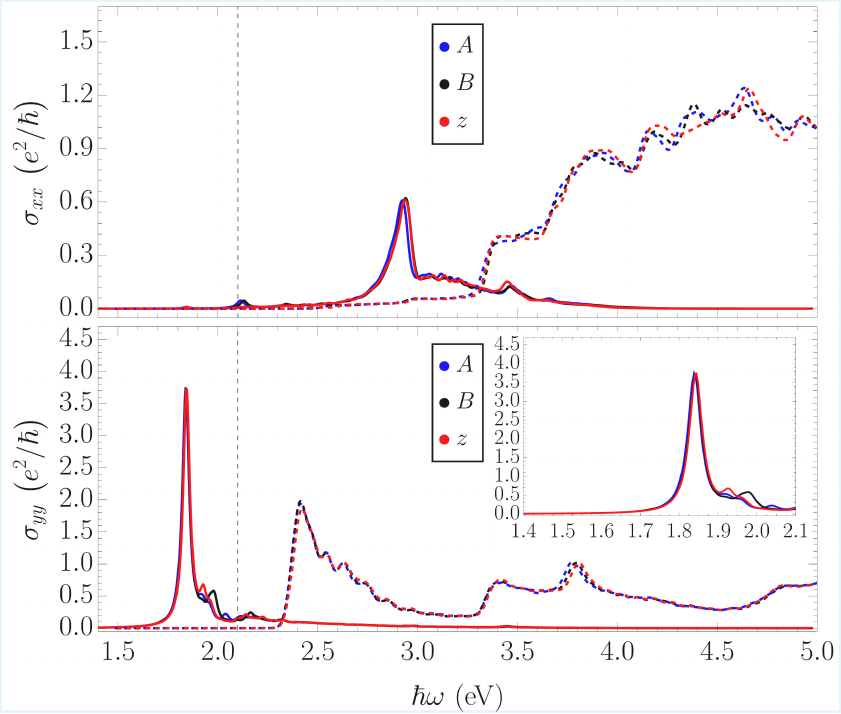}\includegraphics[scale=0.64]{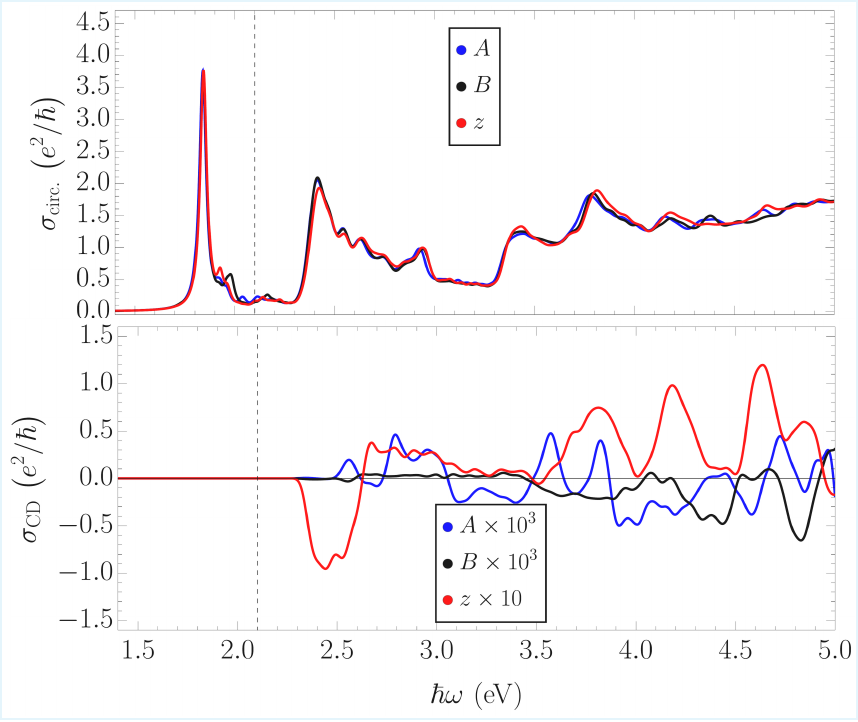}
		\vspace{-0.5cm}\centering\caption{Influence of magnetization direction on linear optical conductivity with both excitonic and IPA contributions for (left panels) linearly polarized light and (right panels) circularly polarized light. Blue, black, and red lines correspond to magnetization chosen to be along the $A$, $B$, and $z$ axis for the Wannierization process, respectively. In all panels, vertical dashed line at $\hbar\omega=2.1\,\mathrm{eV}$ corresponds to the bandgap at the $\Gamma$ point for the unstrained system. For linearly polarized light, excitonic (solid lines) and IPA (dashed lines) contribution to the diagonal tensor elements of the linear optical conductivity (top: $\sigma_{xx}$, bottom: $\sigma_{yy}$) are considered separately as to better showcase the different effects of the magnetization direction on either component. Inset in the bottom-left panel presents the sub-gap region of the conductivity as to more clearly observe the changes in the excitonic peaks.\label{fig:MCD_ABz_difference}}
	\end{figure*}
	
	
	In this work, we investigate the influence of strain on the linear optical response of monolayer \ce{CrSBr}. By solving the Bethe-Salpeter equation, we compute the linear optical conductivity with explicit consideration of excitonic effects. We find that bound excitons predominantly form along the $B$ direction and exhibit pronounced sensitivity to the strain configuration.
	
	Despite the quasi-one-dimensional nature of the exciton wavefunctions, strain applied along both the $A$ and $B$ directions significantly alters their energies and directly impacts the sub-bandgap optical conductivity. When strain is applied collinearly to the polarization direction of the optical conductivity, we observe an energy redshift and blueshift for tensile and compressive strain, respectively. In contrast, strain applied perpendicular to the polarization direction leads to substantial modifications in the spectral lineshape, which we attribute to changes in the energetic ordering of the excitonic series.
	
	Furthermore, magnetic circular dichroism is found to be negligible for bound excitons, with the signal primarily emerging at energies above the bandgap. The MCD signal above the gap is notably sensitive to the orientation of the magnetization.
	This work opens the door to investigate coherent exciton migration controlled by strain \cite{Malakhov2024,quintela2025} in monolayer \ce{CrSBr} and the manipulation of the magnetization at the attosecond time scale by using strong laser pulses \cite{neufeld_attosecond_2023}. For that, we could extend the methodology used here to real-time approaches that also consider excitonic effects \cite{Cistaro_2023,Mosquera_2024}.

	\begin{figure*}
		\centering\includegraphics[scale=0.23]{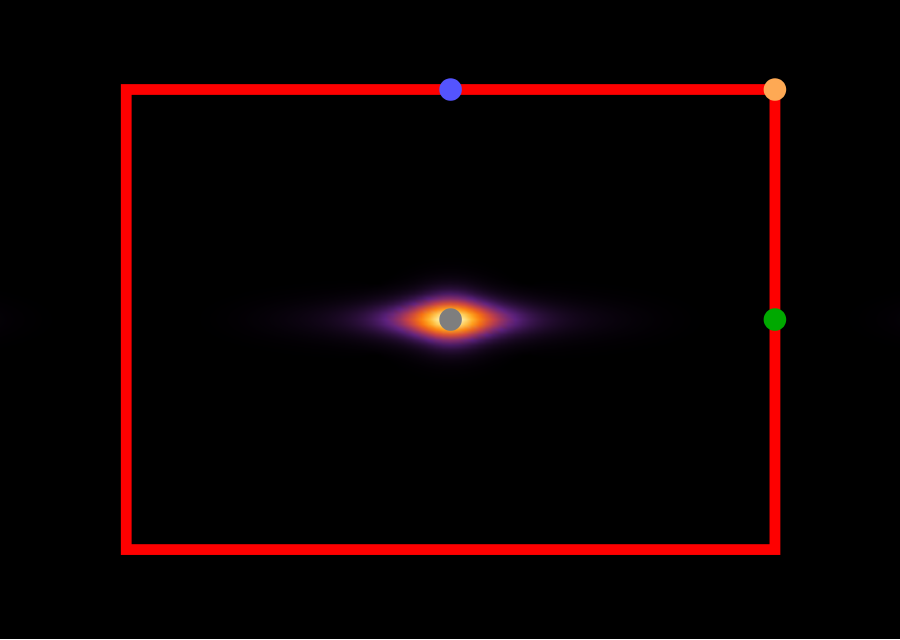}\includegraphics[scale=0.23]{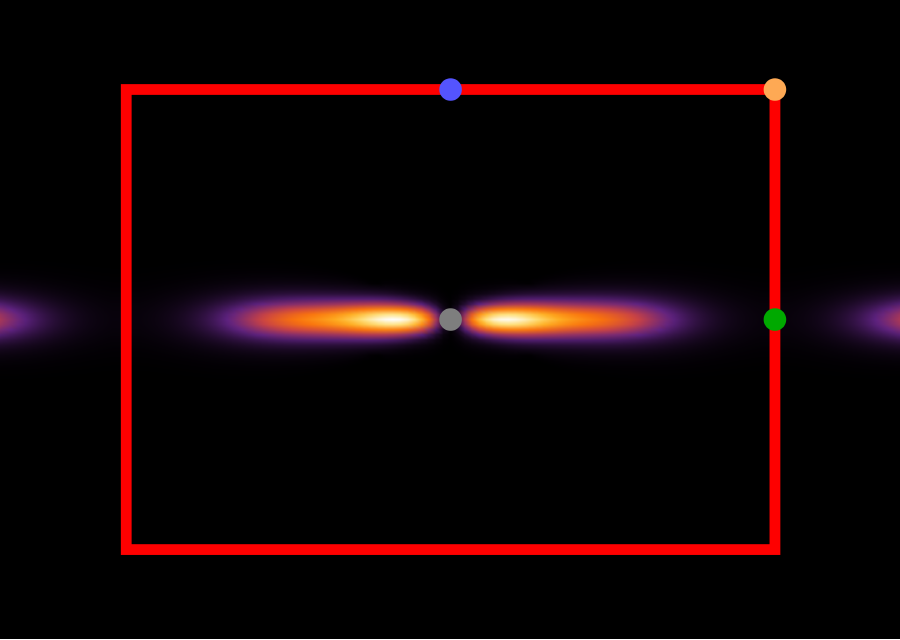}\includegraphics[scale=0.23]{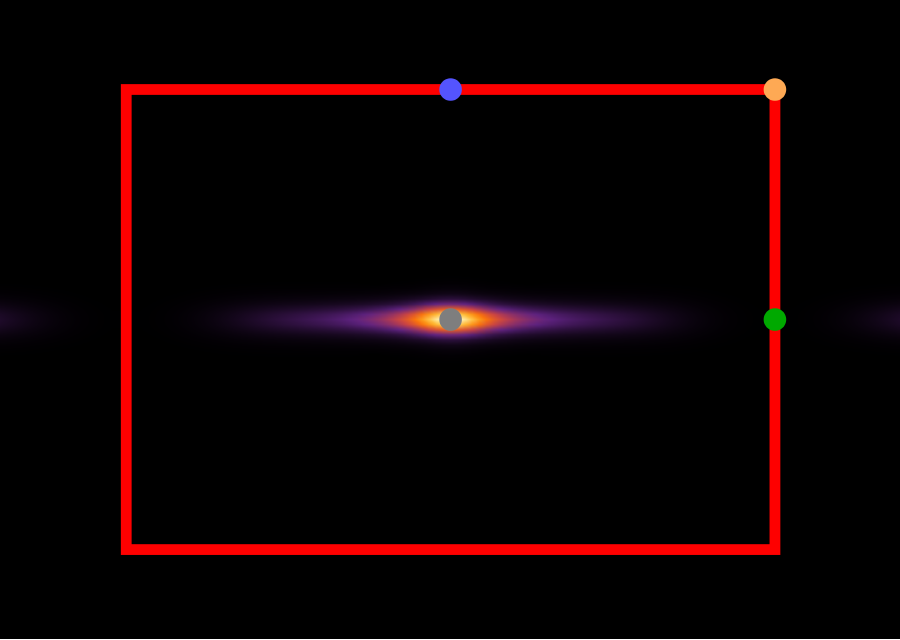}\includegraphics[scale=0.23]{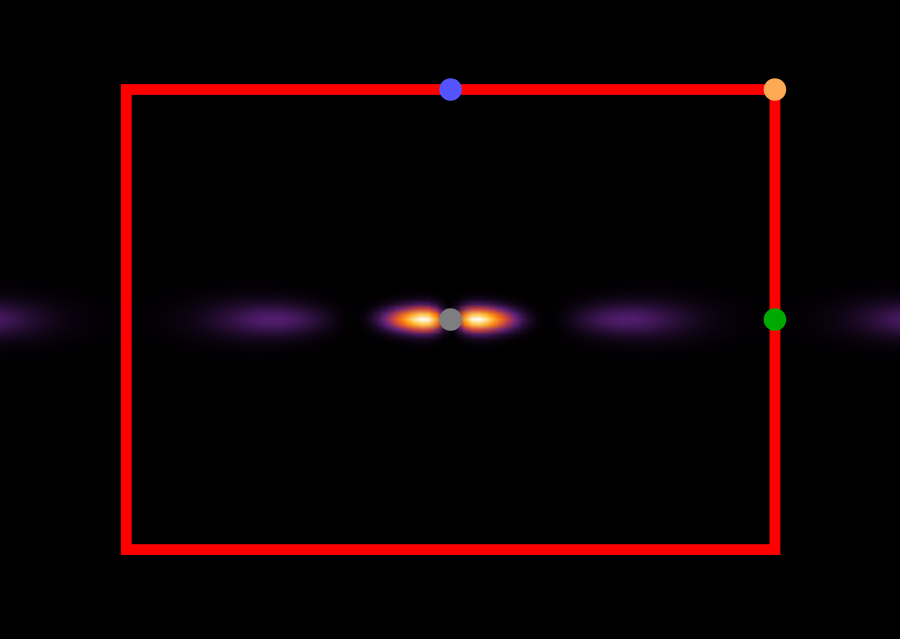}\includegraphics[scale=0.23]{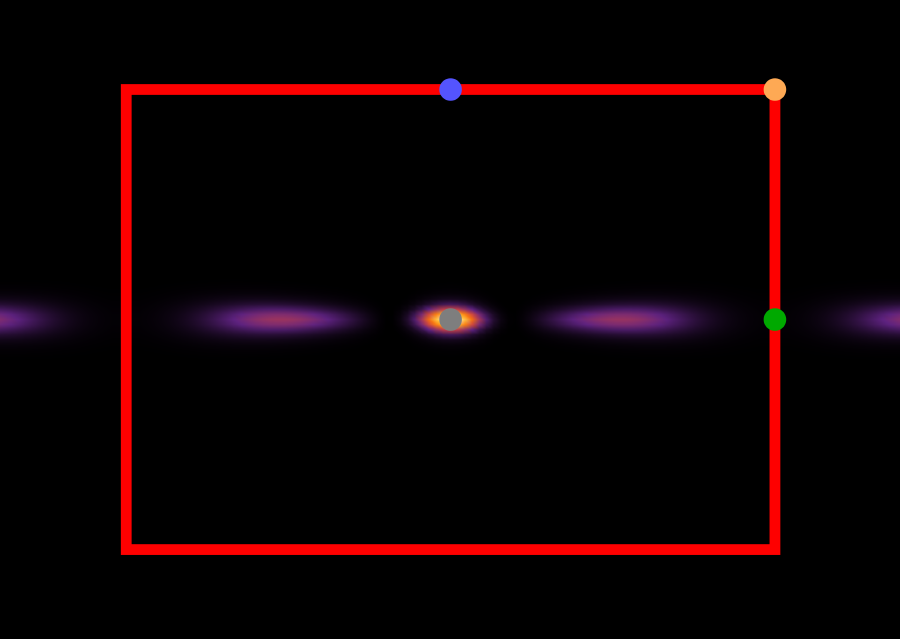}
		\includegraphics[scale=0.64]{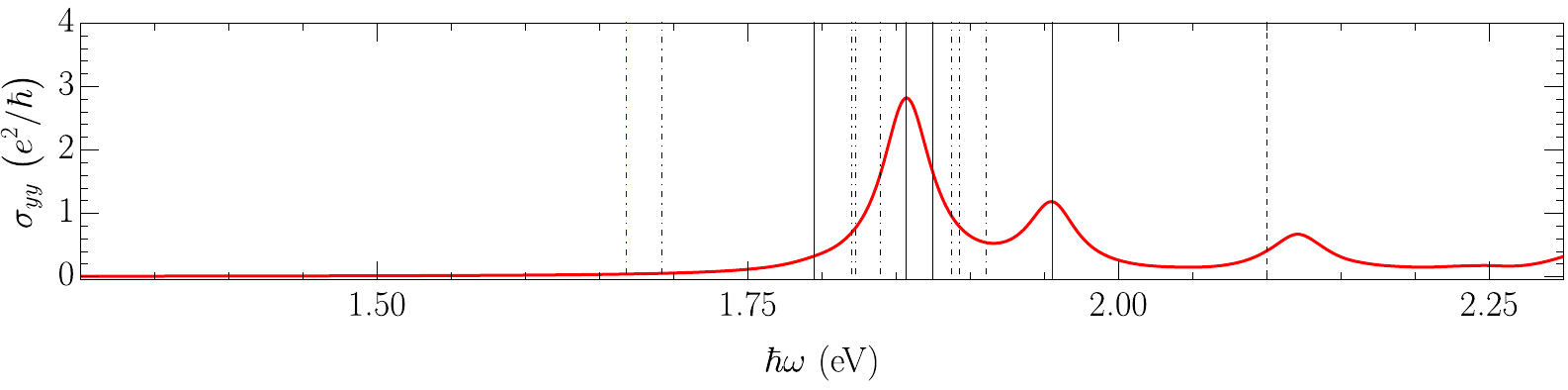}
		
		\vspace{-1.15cm}\centering\includegraphics[scale=0.23]{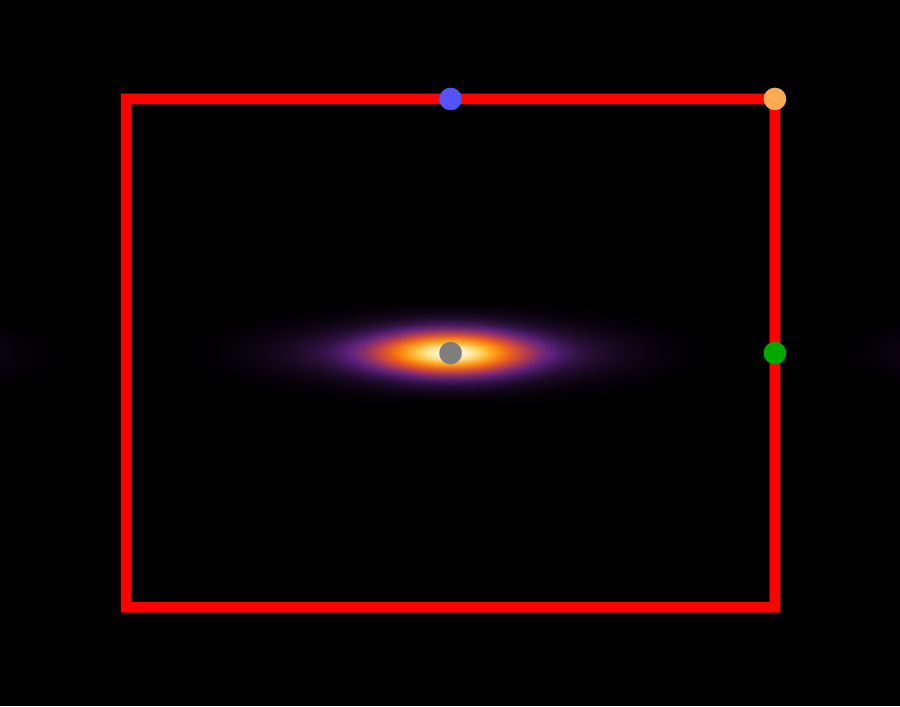}\includegraphics[scale=0.23]{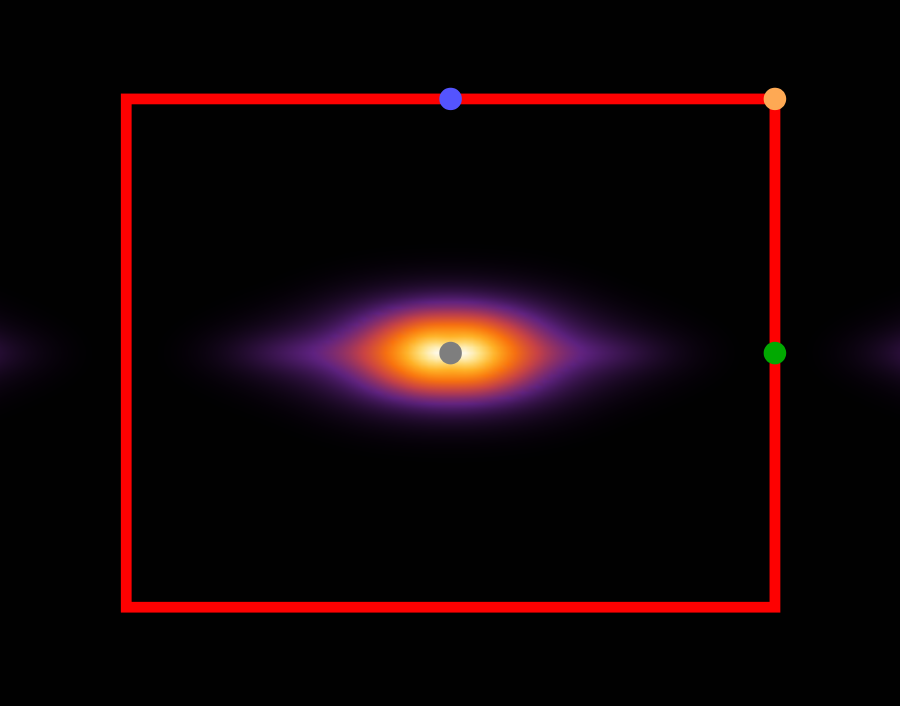}\includegraphics[scale=0.23]{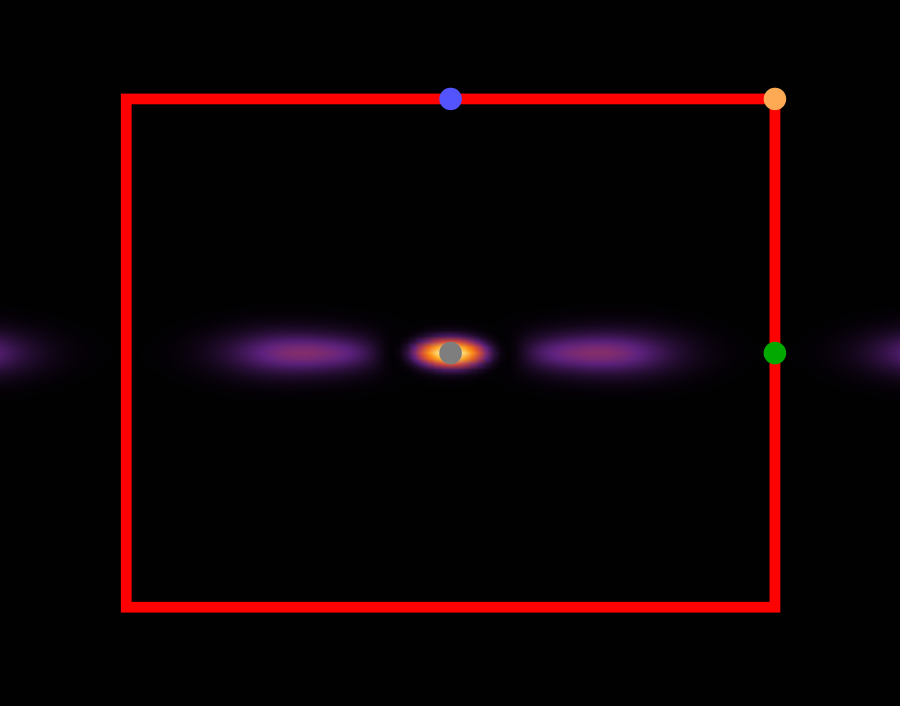}\includegraphics[scale=0.23]{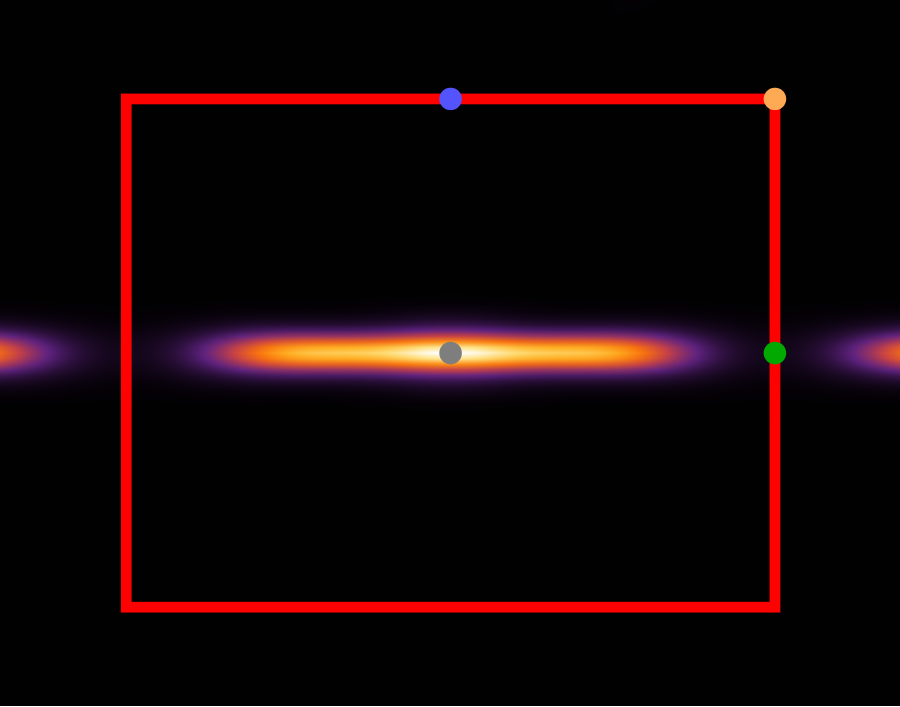}\includegraphics[scale=0.23]{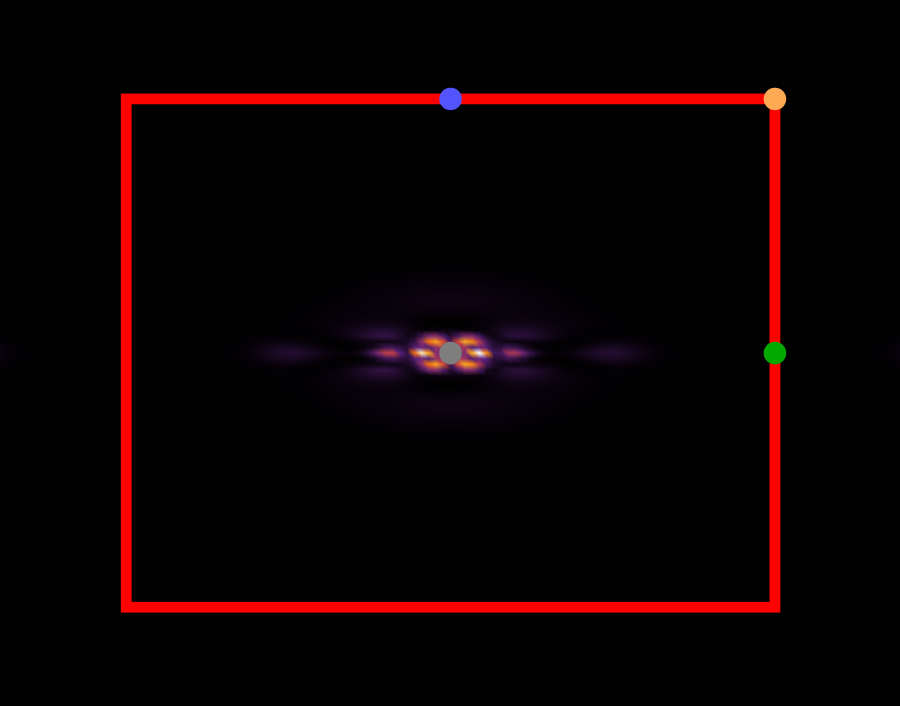}
		\includegraphics[scale=0.64]{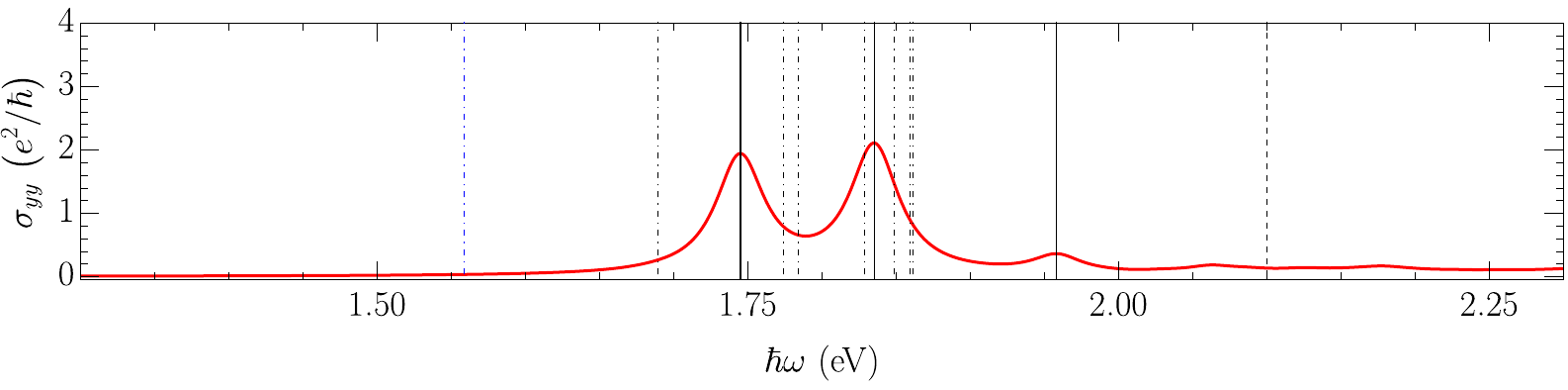}
		
		\vspace{-1.15cm}\centering\includegraphics[scale=0.23]{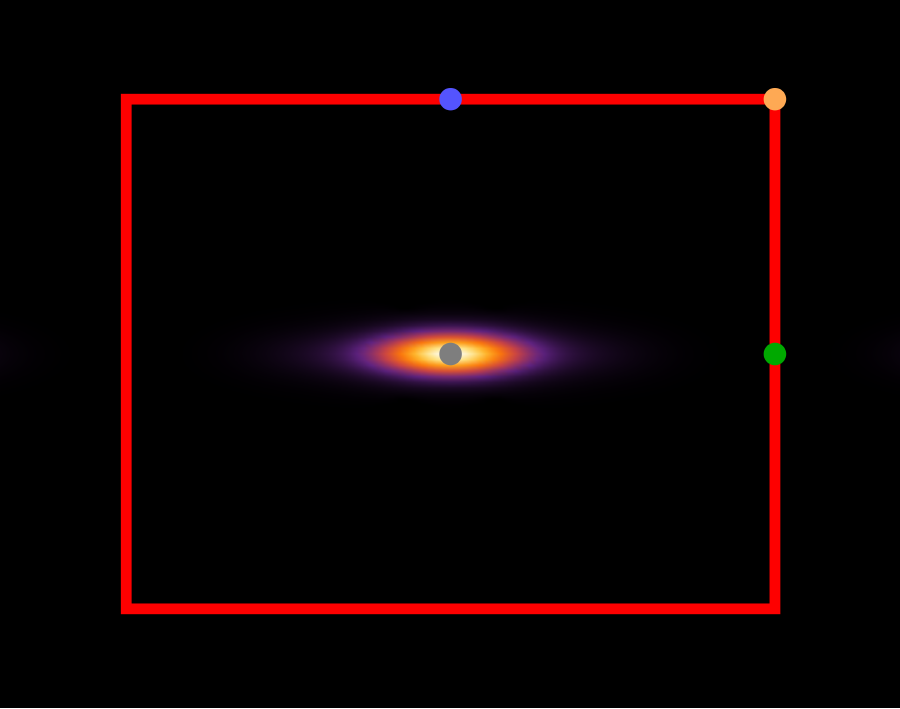}\includegraphics[scale=0.23]{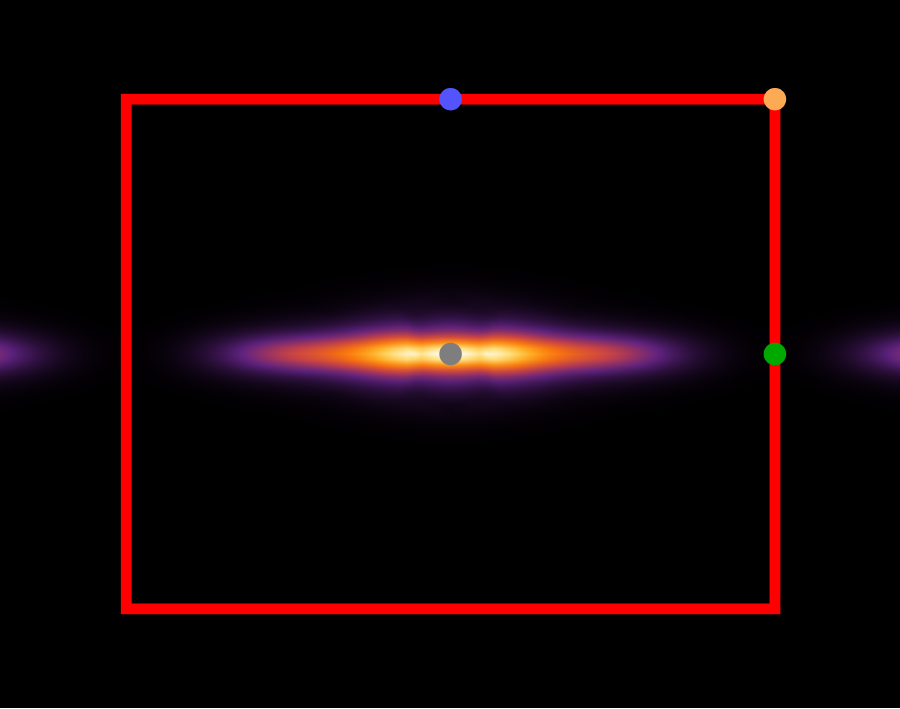}\includegraphics[scale=0.23]{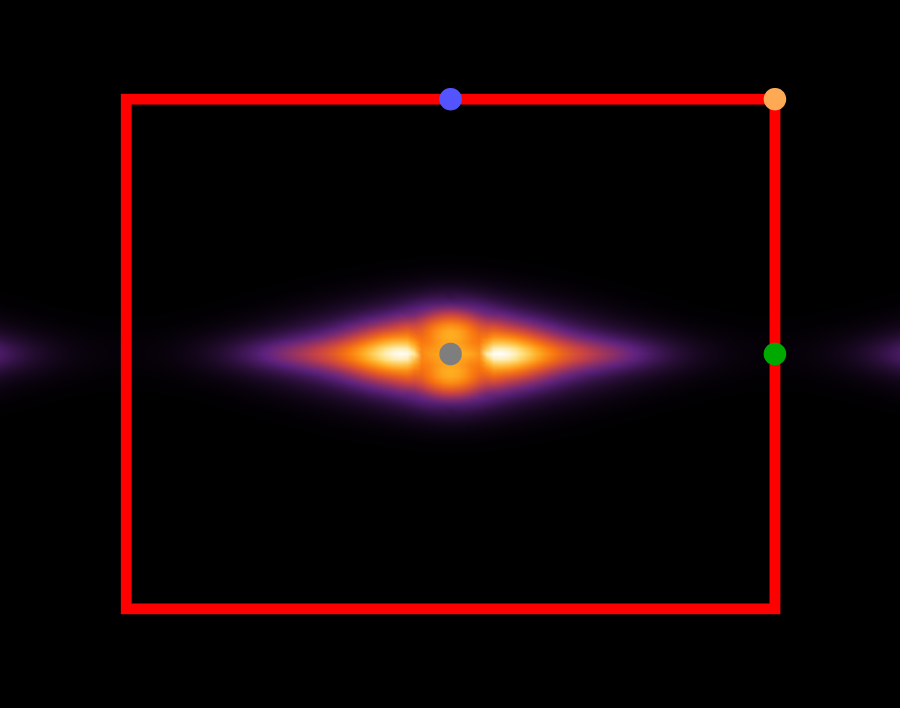}\includegraphics[scale=0.23]{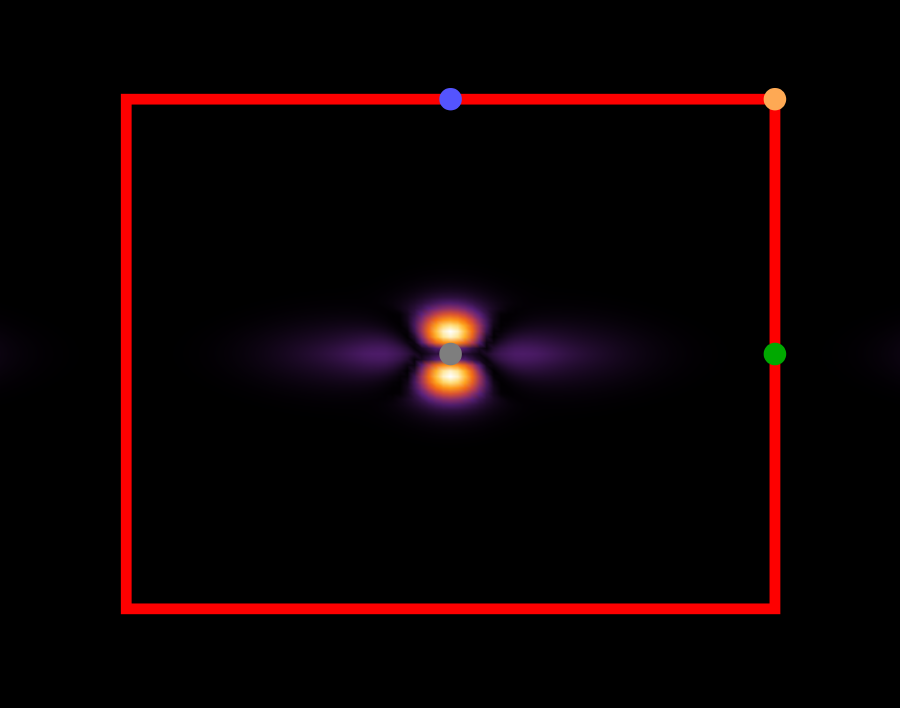}\includegraphics[scale=0.23]{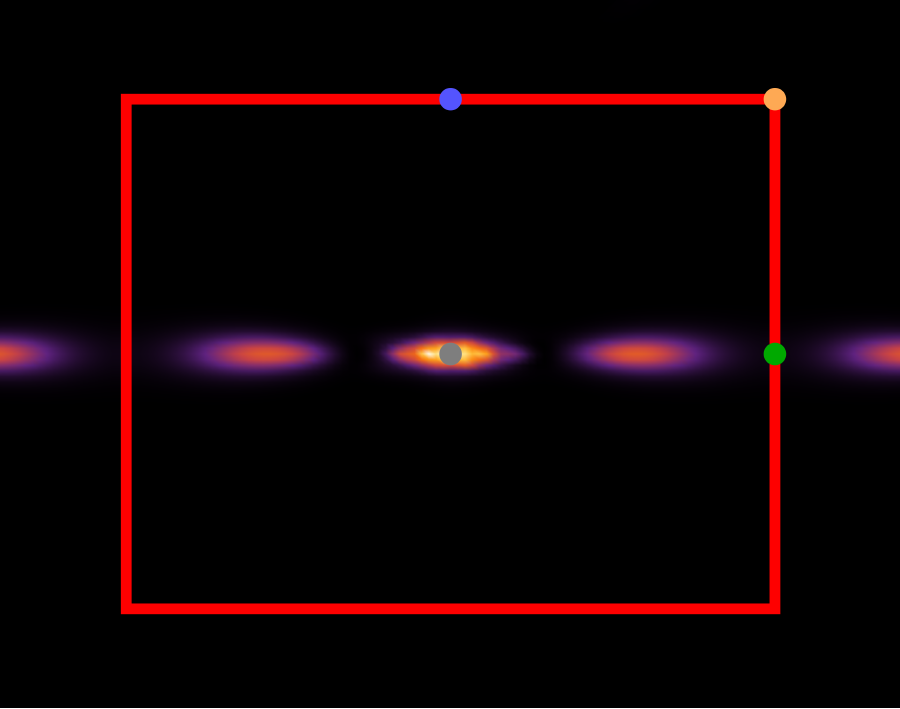}
		\includegraphics[scale=0.64]{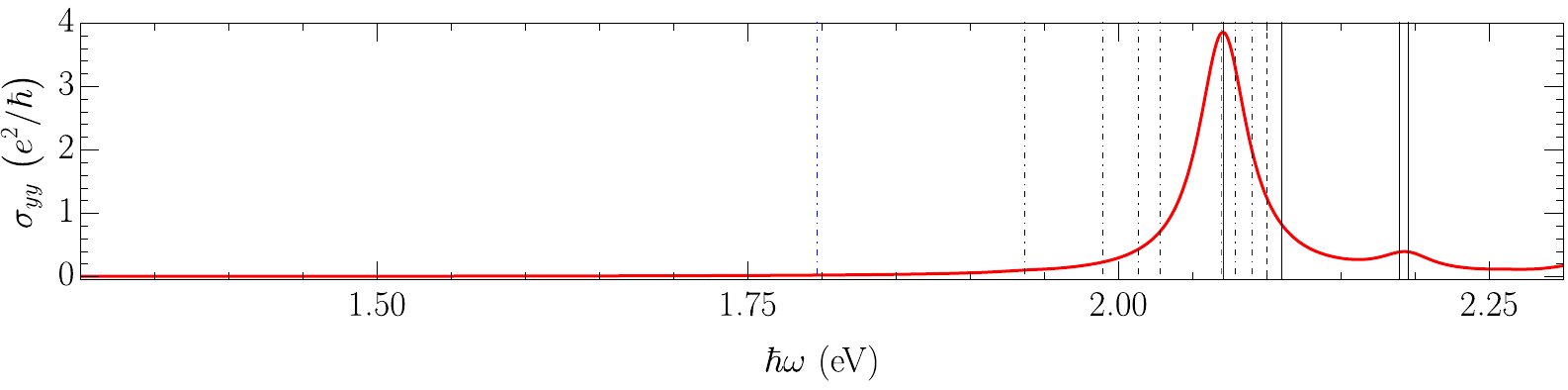}
		
		\vspace{-1.15cm}\centering\includegraphics[scale=0.23]{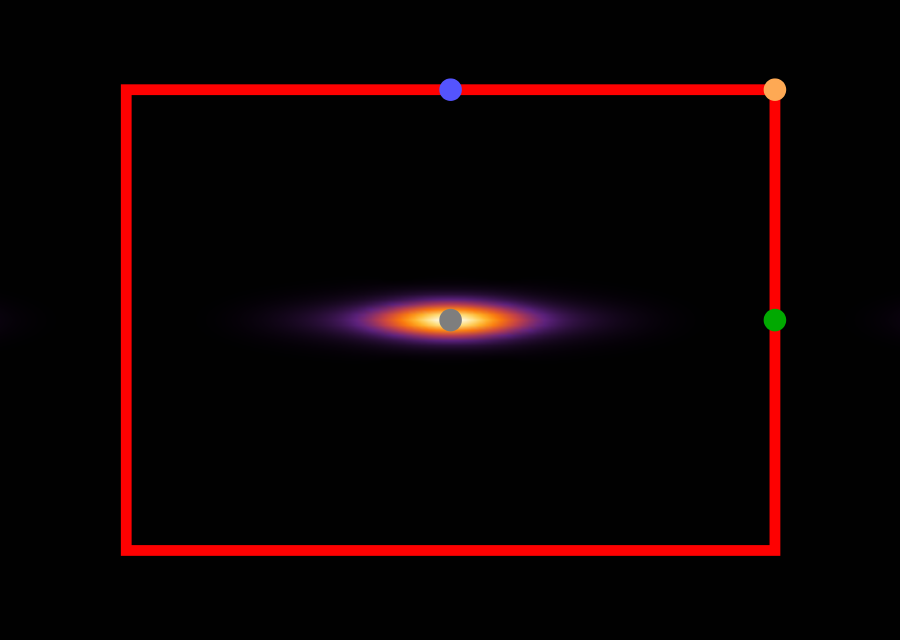}\includegraphics[scale=0.23]{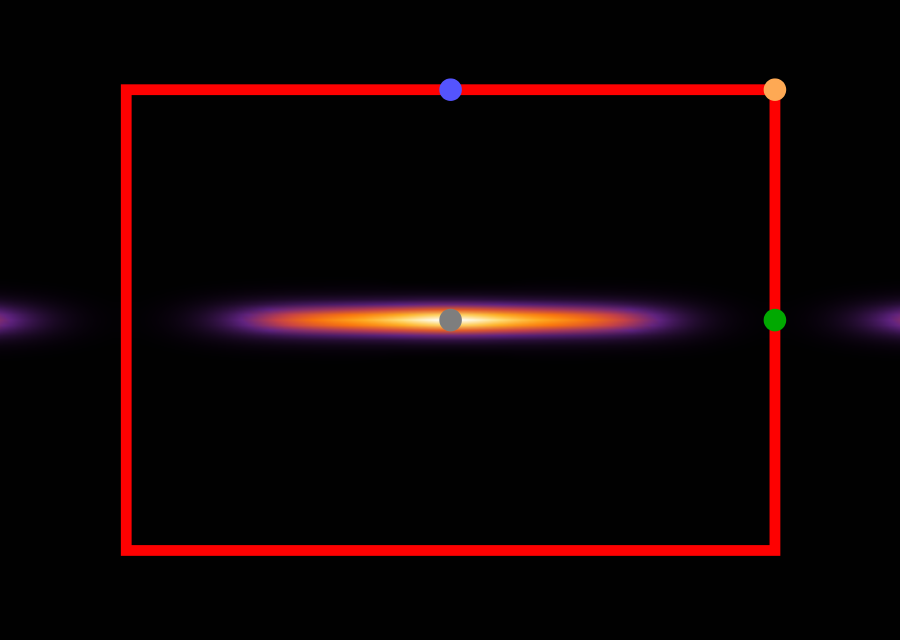}\includegraphics[scale=0.23]{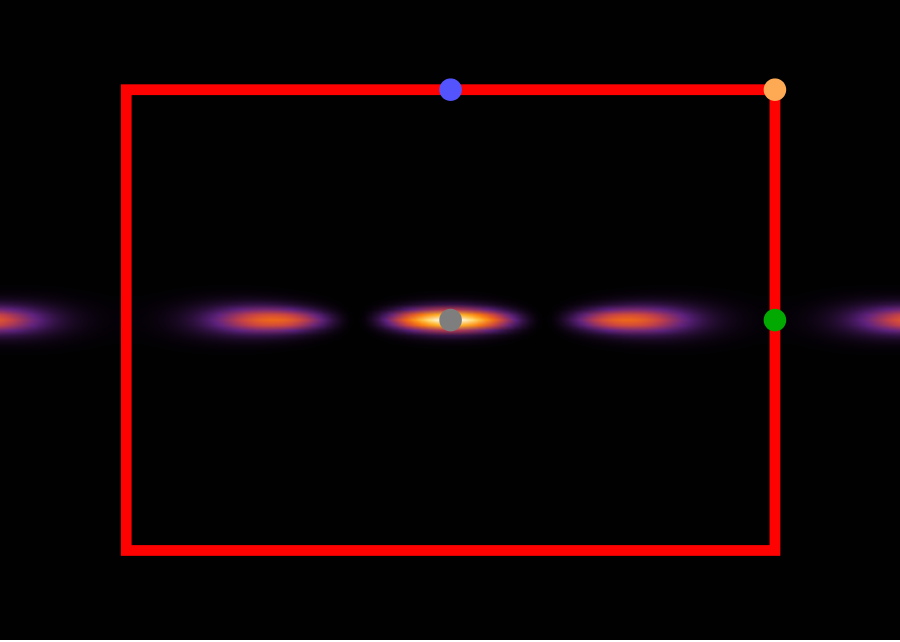}\includegraphics[scale=0.23]{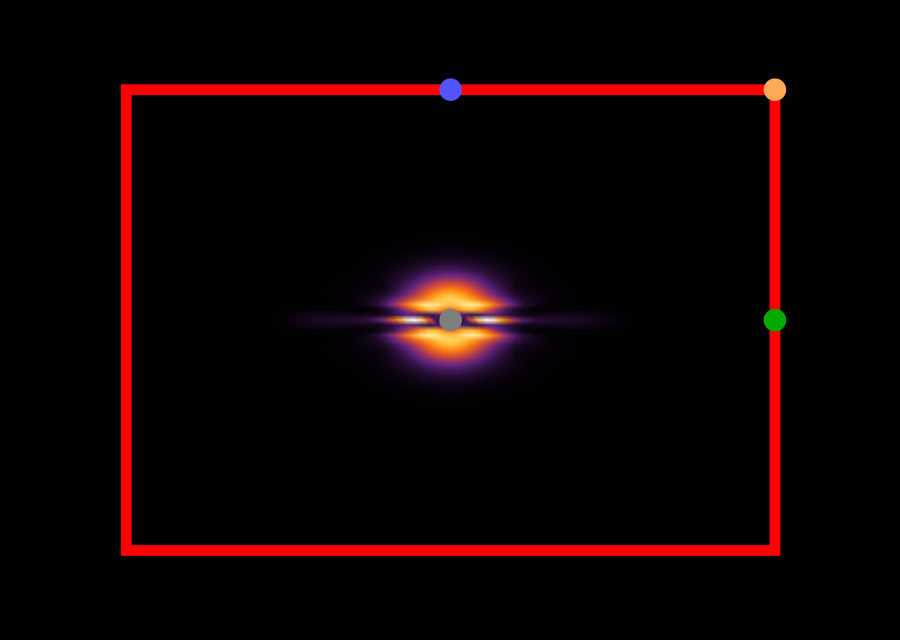}\includegraphics[scale=0.23]{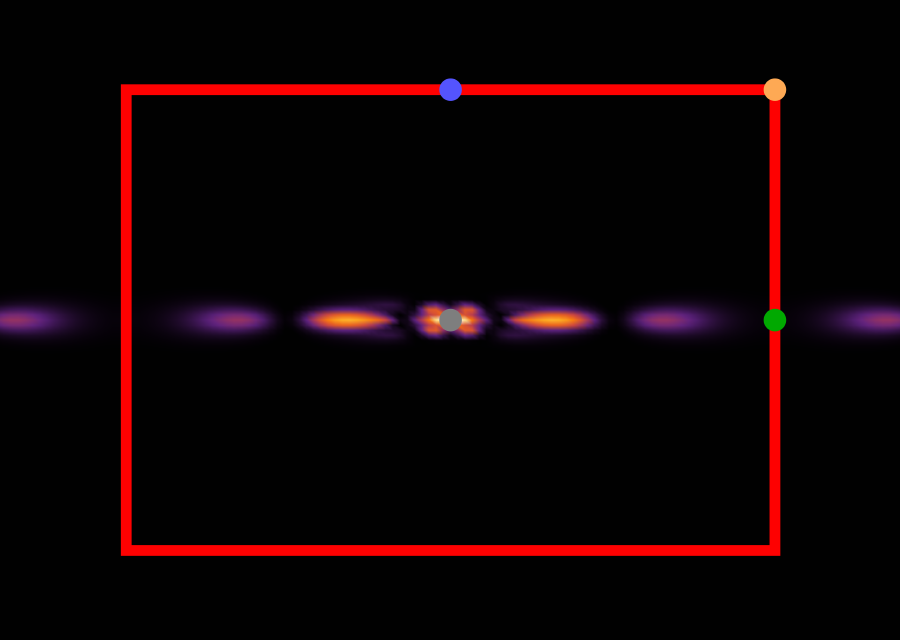}
		\includegraphics[scale=0.64]{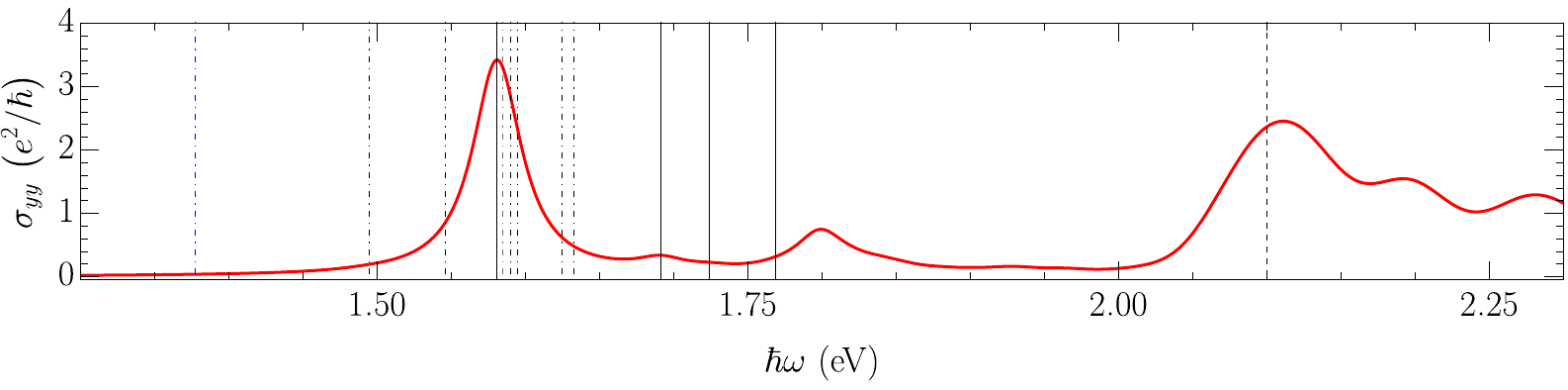}
		\vspace{-0.5cm}\centering\caption{Excitonic wave functions of first (dark) state, together with first four bright states and the corresponding sub-gap response for clearer identification for systems under (top to bottom, in pairs) $95\%\,A$, $105\%\,A$, $95\%\,B$, and $105\%\,B$ strain. \label{fig:wave-function_strain}}
	\end{figure*}
	
	
	\section*{Acknowledgments}
	The authors acknowledge financial support to the Spanish Ministry of Science, Innovation and Universities \& the State Research Agency through grants PID2021-126560NB-I00, CNS2022-135803, TED2021-131323B-I00, and PID2022-141712NB-C21 (MCIU/AEI/FEDER, UE), and the "Mar\'ia de Maeztu" and "Severo Ochoa" Programme for Units of Excellence in R\&D (CEX2023-001316-M and CEX2024-001445-S), "Disruptive 2D materials” (MAD2D-CM-UAM7), funded by Comunidad de Madrid within the Recovery, Transformation and Resilience Plan, and by NextGenerationEU programme from the European Union, the Generalitat Valenciana through the Program Prometeo (2021/017). We also acknowledge computer resources and assistance provided by Centro de Computaci\'on Cient\'ifica de la Universidad Aut\'onoma de Madrid and RES resources (FI-2024-3-0010, FI-2024-3-0011, FI-2024-2-0034, FI-2023-2-0012, FI-2022-3-0022, FI-2022-1-0031). J.J.B. acknowledges the European Union (ERC-2021-StG101042680 2D-SMARTiES) and the Generalitat Valenciana (grant CIDEXG/2023/1), CEX2024-001467-M (ICMol). A.M.R. thanks the Spanish MIU (Grant No FPU21/04195).
	
	\begin{appendices}
		\section{\label{sec:appendix1}Effects of magnetic momentum direction}

In this appendix, we provide a comparison for the $xx$ and $yy$ components of the linear optical conductivity for the three distinct directions of the magnetization chosen during the Wannierization process (specifically, along either $A$ or $B$ axis, or perpendicular to the plane of the monolayer). As can be seen in Fig. \ref{fig:MCD_ABz_difference}, the $xx$ component of the linear optical conductivity barely changes depending on the magnetization choice. On the other hand, the $yy$ component presents quite clear changes in the sub-gap response of the material (see inset in the bottom panel), with the two most pronounced excitonic peaks having quite distinct intensities and locations depending on the axis chosen for the magnetization. However, as can be seen in both the top and bottom panels, the conductivity above the bandgap remains essentially unchanged between the three distinct magnetization directions.

		\section{\label{sec:appendix2}Strain effects on excitonic wave functions}

In this appendix, we provide a comparison between the excitonic states for strains along the $A$ and $B$ directions. These are presented in Fig. \ref{fig:wave-function_strain} for systems under (top to bottom, in pairs) $95\%\,A$, $105\%\,A$, $95\%\,B$, and $105\%\,B$ strain. In the conductivity panels, the lines follow the same notation as those in Fig. \ref{fig:wave-functions}, with blue dot-dashed line corresponding to the lowest energy excitonic state, dot-dashed lines representing first few dark states, solid lines representing the plotted bright states, and dashed line representing the gap at the $\Gamma$ point. As it can be clearly seen, the overall shape of the state corresponding to the dominant peak changes slightly with strain (see third wave function for strain $95\%\,A$ and second wave function for the remaining strains in Fig. \ref{fig:wave-function_strain}), although its position in the excitonic series differs depending on the strain configuration.

	\end{appendices}
	
	\bibliographystyle{ieeetr}
	\bibliography{2D_magnetic}
\end{document}